\pdfoutput=1
\documentclass[11pt,twoside,a4paper,cmspaper,final,collab]{cms-tdr}
\def\svnVersion{bce5e77}\def\svnDate{2022/04/09}\def\cmsCernNoTag{CERN-EP-2021-168}\def\cmsCernDate{\today}\def\cmsMessage{Published in the Journal of High Energy Physics as \href{http://dx.doi.org/10.1007/JHEP04(2022)091}{\doi{10.1007/JHEP04(2022)091}.}}
\begin{document}\cmsNoteHeader{SUS-18-004}

\newlength\cmsTabSkip\setlength{\cmsTabSkip}{1ex}
\providecommand{\cmsTable}[1]{\resizebox{\textwidth}{!}{#1}}
\newcommand{\TChiWZ}{\textsc{TChiWZ}\xspace}
\newcommand{\Ttbffc}{\textsc{T2b}ff\PSGczDo\xspace}
\newcommand{\mczt}{\ensuremath{\widetilde{m}_{\PSGczDt}}\xspace}
\newcommand{\mczo}{\ensuremath{\widetilde{m}_{\PSGczDo}}\xspace}
\newcommand{\mll}{\ensuremath{M(\Pell\Pell)}\xspace}
\newcommand{\msfosll}{\ensuremath{M_{\text{SFOS}}(\Pell\Pell)}\xspace}
\newcommand{\mmnsfosll}{\ensuremath{M^{\text{min}}_{\text{SFOS}}(\Pell\Pell)}\xspace}
\newcommand{\mtlm}{\ensuremath{\mT(\Pell_i,\ptmiss)}\xspace}

\cmsNoteHeader{SUS-18-004}
\title{Search for supersymmetry in final states with two or three soft leptons and missing transverse momentum in proton-proton collisions at \texorpdfstring{$\sqrt{s} = 13\TeV$}{sqrt(s) = 13 TeV}}

\date{\today}

\abstract{
    A search for supersymmetry in events with two or three low-momentum leptons and missing transverse momentum is performed. The search uses proton-proton collisions at $\sqrt{s} = 13\TeV$ collected in the three-year period 2016--2018 by the CMS experiment at the LHC and corresponding to an integrated luminosity of up to 137\fbinv. The data are found to be in agreement with expectations from standard model processes. The results are interpreted in terms of electroweakino and top squark pair production with a small mass difference between the produced supersymmetric particles and the lightest neutralino.
    For the electroweakino interpretation, two simplified models are used, a wino-bino model and a higgsino model. Exclusion limits at 95\% confidence level are set on \PSGczDt/\PSGcpmDo masses up to 275\GeV for a mass difference of 10\GeV in the wino-bino case, and up to 205(150)\GeV for a mass difference of 7.5\,(3)\GeV in the higgsino case. The results for the higgsino are further interpreted using a phenomenological minimal supersymmetric standard model, excluding the higgsino mass parameter $\mu$ up to 180\GeV with the bino mass parameter $M_1$ at 800\GeV. In the top squark interpretation, exclusion limits are set at top squark masses up to 540\GeV for four-body top squark decays and up to 480\GeV for chargino-mediated decays with a mass difference of 30\GeV.
}

\hypersetup{
pdfauthor={CMS Collaboration},
pdftitle={Search for supersymmetry in final states with two or three soft leptons and missing transverse momentum in proton-proton collisions at sqrt(s)=13 TeV},
pdfsubject={CMS},
pdfkeywords={CMS, SUSY, compressed, leptons, missing energy, 13 TeV}}

\maketitle
\section{Introduction}\label{sec:introduction}

Numerous searches for physics beyond the standard model (SM) at the CERN LHC target weakly interacting massive particles (WIMPs) as potential candidates to explain the nature of dark matter (DM). These WIMPs are predicted to be produced either directly through the interaction of SM particles or indirectly through the decays of new particles with higher mass. Stable WIMPs would escape undetected, yielding a characteristic signature of significant missing transverse momentum (\ptmiss) together with any energetic leptons, photons and hadronic jets produced in the cascade of the decaying higher-mass states. In scenarios where the WIMP is the lightest among new particles with a nearly degenerate mass spectrum, the production of these new particles leads to events with relatively low visible energy, and by inference, also low \ptmiss. These scenarios are referred to as compressed mass spectra.
 
Among theories beyond the SM, Supersymmetry (SUSY) \cite{SUSY0,SUSY1,SUSY2,SUSY3,SUSY4} is particularly attractive owing to the fact that it can provide answers  to questions unanswered in the context of the SM, such as the naturalness of the theory \cite{LightStop1,Witten:1981nf,Dimopoulos:1981zb} and the nature of DM. In models where $R$-parity conservation is imposed~\cite{Farrar:1978xj}, SUSY particles can only be produced in pairs, and each of their decay chains must lead to the lightest SUSY particle (LSP), which has to be stable. In many models this corresponds to the lightest neutralino (\PSGczDo), a neutral and stable WIMP, which is thus a viable DM candidate~\cite{PDG2020}. 
As indicated in several studies, such as in Refs.~\cite{LightStop1,LightStop2,Dimopoulos:1981zb,Witten:1981nf,Dine:1981za,Dimopoulos:1981au,Sakai:1981gr,Kaul:1981hi}, naturalness imposes constraints on the masses of higgsinos, top squarks, and gluinos, placing them potentially within the reach of the experiments at the LHC.

Despite numerous searches for new particles, no experimental evidence for their production in LHC collisions has been found, and strong experimental constraints on the masses of SUSY particles have been set. As a result, the relevance of SUSY at the electroweak scale has come under intense scrutiny and several possibilities of how electroweak-scale SUSY might have escaped detection are being considered.
Compressed mass spectra offer a possible scenario in which low-energy SUSY might have eluded experiments thus far, if the new particles exist in a region of SUSY parameter space where experimental searches have lower sensitivity. 

In compressed spectra scenarios, most of the energy and momentum of higher-mass states is carried away in the rest energy of the LSPs and the remaining detectable SM particles have low momentum. Events with such characteristics can be distinguished from bulk SM processes by requiring a jet with large transverse momentum (\pt) from initial state radiation (ISR) that leads to a large boost of the SUSY particle pair and thus large \ptmiss. Final states with only soft jets and moderate \ptmiss do not exhibit high sensitivity to new physics because of the presence of huge backgrounds from quantum chromodynamics (QCD) multijet production and $\PZ$+jets events with invisible \PZ boson decays. In the latter case, the SM background can be reduced effectively with the additional requirement of two or three soft light leptons (electrons or muons).

These soft leptons, along with moderate or large \ptmiss, constitute the signature of the search presented in this paper.  Specifically, the production of charginos (\PSGcpmDo) and neutralinos (\PSGczDo) with nearly degenerate mass, decaying to soft leptons and \ptmiss is considered. In what follows, neutralinos and charginos are collectively referred to as electroweakinos.

The results are interpreted in the context of $R$-parity conserving SUSY.
A simplified wino-bino model is probed. In such scenarios the LSP can be a WIMP DM candidate that was depleted in the early universe through co-annihilation processes to match the observed DM density \cite{Griest:1990kh,Edsjo:1997bg}. 
Light higgsinos, which are favored by naturalness arguments, are likely to be nearly degenerate in mass~\cite{Baer:2011ec,Baer:2014kya,Han:2013usa,Han:2014kaa}. Two minimal SUSY SM (MSSM) hypotheses, where the lightest electroweakinos are higgsino-like, are tested: a simplified higgsino model and a phenomenological MSSM higgsino model based on the pMSSM~\cite{Djouadi:1998di}.
The search is also sensitive to top squark pair production models where a light top squark and the LSP are nearly degenerate in mass and the top squark decays directly to four fermions. The near-degeneracy in mass of the top squark and the LSP is a typical example of the so-called co-annihilation region, in which the LSP is the sole source of DM~\cite{Coannihilation}.

Results of searches in final states with soft leptons and missing transverse momentum were previously presented by ATLAS and CMS using data sets at center of mass energy 8 and 13\TeV \cite{CMS:2015epp, sos2016,ATLAS:2017vat,PhysRevD.101.052005,ATLASTrilepton}. The previous iteration of this CMS analysis targeted signal events with moderate or large \ptmiss, by requiring an ISR jet and a pair of soft oppositely charged (opposite sign (OS)) leptons.
The analysis presented in this paper extends the previous search with the addition of soft trilepton final states, as well as a relaxed selection on the dilepton invariant mass (from 4 to 1\GeV) and reoptimized signal regions, yielding extended sensitivity towards lower values of the \PSGczDt-\PSGczDo mass difference ($\Delta m$), where \PSGczDt is the next to lightest neutralino. The analysis also employs improved methods for the estimation of the nonprompt lepton background and uses the CMS data set collected during 2016--2018, corresponding to an integrated luminosity of up to 137\fbinv.
The final results are extracted from a simultaneous binned maximum likelihood fit to the data of the signal and background expectations from all the signal and control regions. The experimental and theoretical uncertainties affecting the background and signal estimations are incorporated as nuisance parameters. The results of this analysis are included in tabulated form in the HEPData record~\cite{hepdata}.

The paper is organized as follows: Section~\ref{sec:cms} summarizes the general features of the CMS detector, while Section~\ref{sec:sample} describes the data set and simulation samples used in this search. After defining the physics objects in Section~\ref{sec:objects}, the event selection and the signal regions of the search are described in Section~\ref{sec:eventselection}. The background estimation and the systematic uncertainties are discussed in Sections~\ref{sec:background} and \ref{sec:systematics}, respectively. The results of the search are presented in Section~\ref{sec:results} and their interpretations in the relevant SUSY scenarios in Section~\ref{sec:interpretations}. The summary of the paper is given in Section~\ref{sec:summary}.

\section{The CMS detector}\label{sec:cms}

The central feature of the CMS apparatus is a superconducting solenoid of 6\unit{m} internal diameter, providing a magnetic field of 3.8\unit{T}. Within the solenoid volume are a silicon pixel and strip tracker, a lead tungstate crystal electromagnetic calorimeter (ECAL), and a brass and scintillator hadron calorimeter (HCAL), each composed of a barrel and two endcap sections. Forward calorimeters extend the pseudorapidity ($\eta$) coverage provided by the barrel and endcap detectors. Muons are detected in gas-ionization chambers embedded in the steel flux-return yoke outside the solenoid. Events of interest are selected using a two-tiered trigger system. The first level, composed of custom hardware processors, uses information from the calorimeters and muon detectors to select events at a rate of around 100\unit{kHz} within a fixed latency of about 4\mus~\cite{Sirunyan:2020zal}. The second level, known as the high-level trigger, consists of a farm of processors running a version of the full event reconstruction software optimized for fast processing, and reduces the event rate to around 1\unit{kHz} before data storage~\cite{Khachatryan:2016bia}. 

A more detailed description of the CMS detector, together with a definition of the coordinate system used and the relevant kinematic variables, can be found in Ref.~\cite{Chatrchyan:2008zzk}.

\section{Data and simulated samples}\label{sec:sample}

The data used in this search have been collected by the high-level trigger system using three different algorithms (trigger paths): First, an inclusive trigger path requiring the selected events to have online $\ptmiss > 120\GeV$, where \ptmiss is the missing transverse momentum, corrected to account for the contribution of muons. Second, a trigger path that requires at least two muons in addition to \ptmiss, with a lower online threshold $\ptmiss > 60\GeV$. This trigger path also requires the online raw \ptmiss, \ie not corrected for the presence of muons, to be greater than 50\GeV, the \pt of each muon to satisfy $\pt > 3\GeV$, and the invariant mass of the muon pair to be between 3.8 and 56\GeV. As this last requirement becomes limiting for events used in the $\PW\PZ$-enriched region of the analysis (defined in Section~\ref{sec:background}), a third trigger path, with a requirement only on the \pt of the highest \pt muons ($\pt(\Pell_1) > 17\GeV$ and $\pt(\Pell_2) > 8\GeV$), is also used. Some of the trigger requirements affect the event selection, \eg, the relaxed selection of the dilepton invariant mass. This is discussed in greater detail in Section~\ref{sec:eventselection}.

The data sample collected with the inclusive \ptmiss trigger corresponds to the total integrated luminosity of each year (35.9\fbinv in 2016, 41.5\fbinv in 2017, and 59.7\fbinv in 2018). The luminosity that corresponds to the dimuon-only trigger path is slightly lower in 2017 (36.7\fbinv), while it is slightly lower for the dimuon+\ptmiss trigger path in all years (33.2\fbinv in 2016, and 59.2\fbinv in 2018), due to the fact that the trigger paths were disabled for some runs.

Simulated signal and major background processes, such as \ttbar, DY, $\PW$+jets, and $\PZ$+jets are generated with the \MGvATNLO \cite{Alwall:2014hca,Frederix:2012ps} event generator at leading order (LO) precision in perturbative QCD, with the MLM merging scheme \cite{Alwall:2007fs} used to consolidate additional partons from the event generator with parton shower generator. The diboson processes $\PW\PW$, $\PZ\PZ$, and $\PW\PGg$ are generated with the same event generator as above at next-to-LO (NLO) precision using the {FxFx} merging scheme~\cite{Frederix:2012ps}, while the $\PW\PZ$ process is generated at NLO with {\POWHEG}~v2.0~\cite{Nason:2004rx,Frixione:2007vw,Alioli:2010xd,Melia:2011tj,Nason:2013ydw}. Rare background processes (\eg, $\ttbar\PW$, $\ttbar\PZ$, $\PW\PW\PW$, $\PZ\PZ\PZ$, $\PW\PZ\PZ$, and $\PW\PW\PZ$) are also generated at NLO precision with \MGvATNLO \cite{Alwall:2014hca,Frederix:2012ps}. The rare background from single top quarks produced in association with a \PW boson is generated at NLO precision with {\POWHEG}~v1.0~\cite{Re:2010bp}. The \MGvATNLO versions used are the 2.2.2 (2.3.2 for $\ttbar\PZ$) for 2016 and 2.4.2 (2.6.5 for $\ttbar\PZ$) for 2017 and 2018. The NNPDF3.0~\cite{Ball:2014uwa} (2016) or NNPDF3.1~\cite{Ball2017} (2017 and 2018) LO and NLO parton distribution functions (PDF) are used for the simulated samples generated at LO and NLO respectively. Showering, hadronization, and the underlying event description are carried out using the \PYTHIA 8.212 package \cite{Sjostrand:2014zea} with the CUETP8M1 (CP5) underlying event tune for 2016~\cite{Skands:2014pea,CMS-PAS-GEN-14-001} (2017 and 2018~\cite{CP5tune}). A detailed simulation of the CMS detector, based on the \GEANTfour~\cite{Agostinelli:2002hh} package, is performed.

\begin{figure}[!hbtp]
\centering
\includegraphics[width=0.35\textwidth]{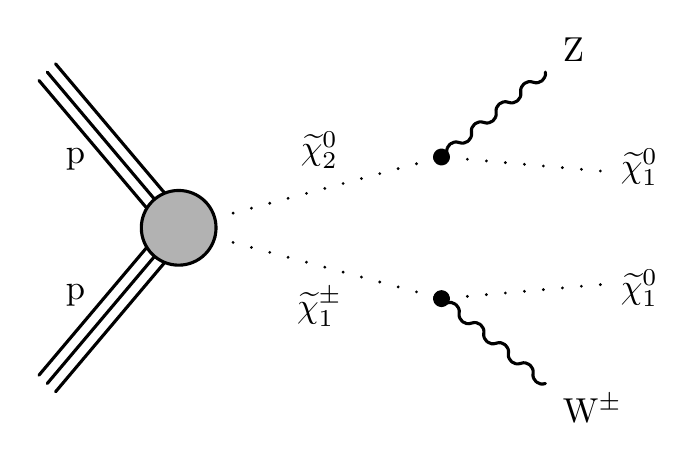}
\includegraphics[width=0.35\textwidth]{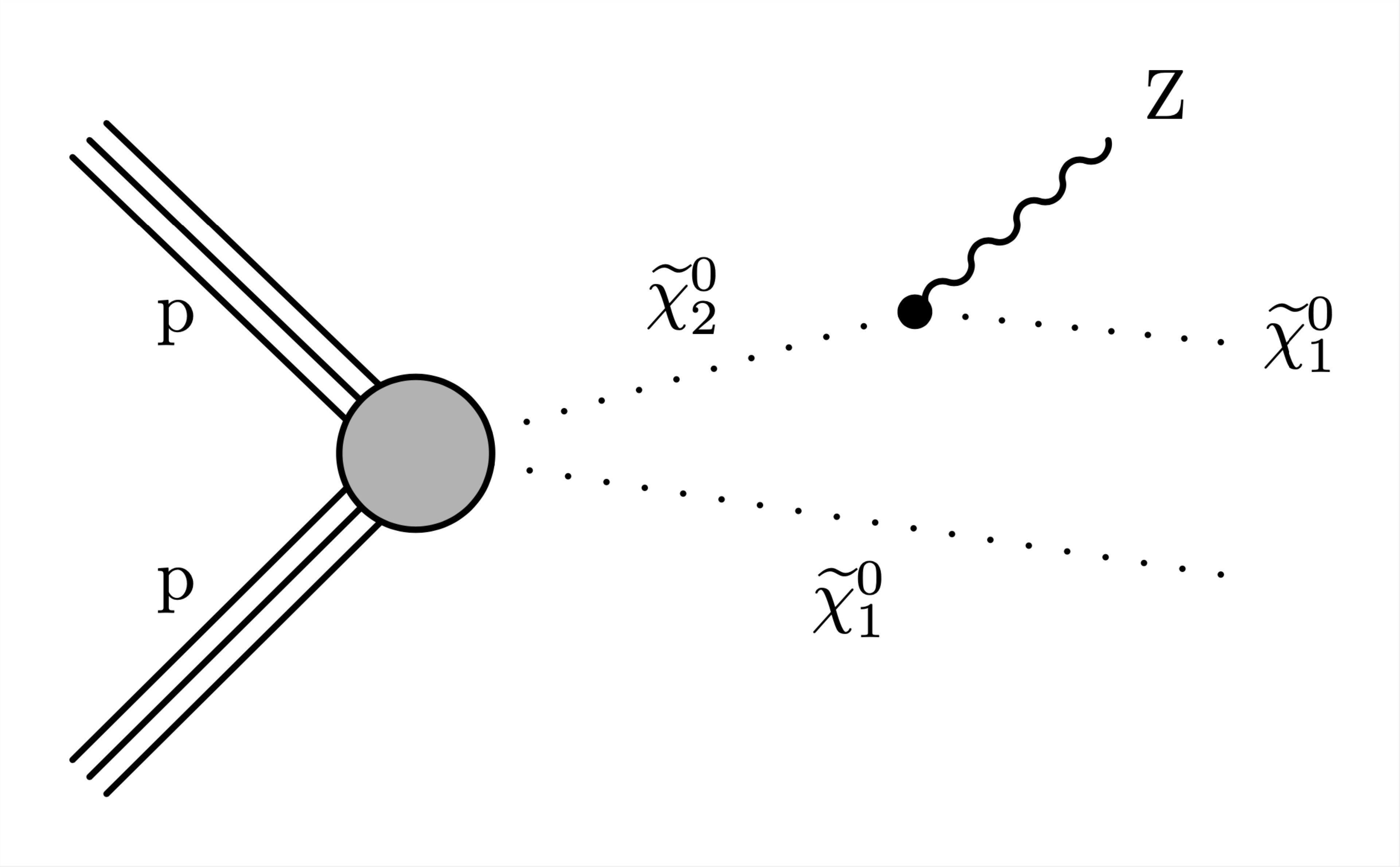}
\includegraphics[width=0.35\textwidth]{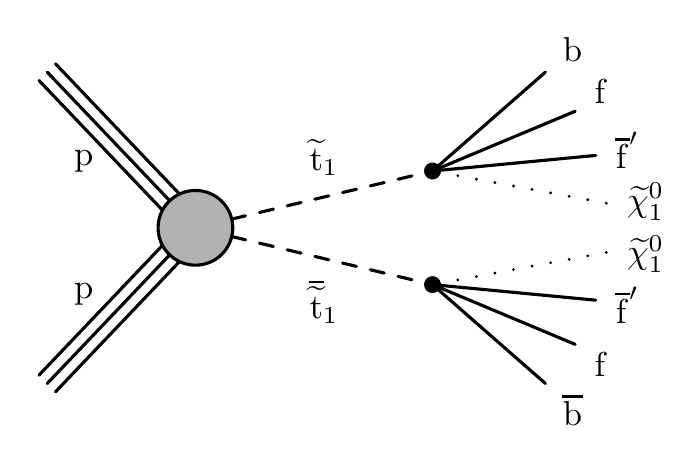}
\includegraphics[width=0.35\textwidth]{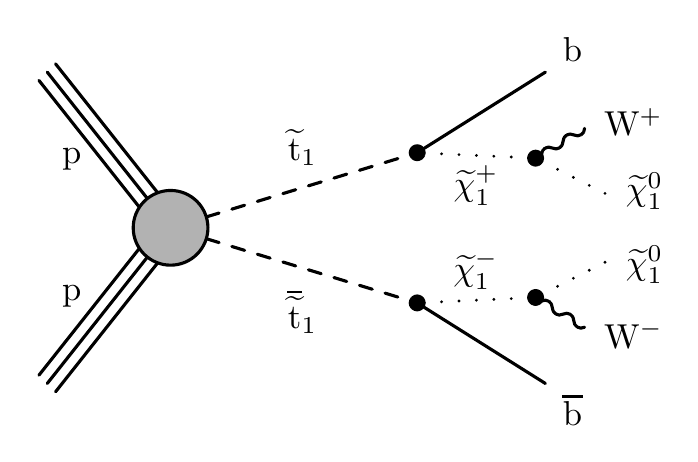}
\caption{Production and decay of electroweakinos in the \TChiWZ model (upper left), in the higgsino simplified model (upper left and right), in the \Ttbffc model (lower left) and in the \textsc{T2bW} model (lower right).}
\label{fig:Signals}
\end{figure}

The five signal models considered in this analysis include the production of electroweakinos and top squark pairs. Simplified models~\cite{Alwall:2008ag, LHCNewPhysicsWorkingGroup:2011mji,CMS:2013wdg} are used in which all SUSY particles other than the electroweakinos (or the top squarks) under study are assumed to be too massive to affect the analysis observables. Such models target scenarios with a bino LSP and wino next-to-LSP (NLSP), a higgsino LSP, and top squarks decaying to electroweakino LSPs. In all of the simplified models in this search, the assumption of 100\% branching fractions to a single, representative decay is made. A pMSSM-inspired model is also considered and is described below. Figure \ref{fig:Signals} shows representative diagrams for simplified models of electroweakino and top squark pair production.

In the wino-bino simplified model (denoted as \TChiWZ), the neutralino-chargino (\PSGczDt-\PSGcpmDo) pair production process is studied assuming a pure wino production cross section, where the $\PSGczDt$ and $\PSGcpmDo$ are assumed to be mass degenerate and taken to decay to the LSP via virtual \PZ\ and \PW\ bosons. For the higgsino simplified model, production through both $\PSGczDt\PSGcpmDo$ and $\PSGczDt\PSGczDo$ channels is considered and $m_{\PSGcpmDo}=\frac{1}{2}(m_{\PSGczDo}+m_{\PSGczDt})$ is assumed. In each case, the production cross sections are computed at NLO plus next-to-leading logarithmic (NLL) precision~\cite{Beenakker:1999xh,Fuks:2012qx,Fuks:2013vua} and correspond to the cases of pure wino and pure higgsino states, respectively. Mass differences between the $\PSGczDt$ and $\PSGczDo$ ranging from 1 to 50\GeV are considered in these simplified models of electroweakino production.

A model inspired by the pMSSM is used for further interpretations in the case of a higgsino LSP. For a higgsino LSP in the pMSSM, the physical mass eigenstates of the $\PSGczDo$, $\PSGcpmDo$, and $\PSGczDt$ are determined primarily by the higgsino, bino, and wino mass parameters ($\mu$, $M_1$, and $M_2$, respectively) through the neutralino and chargino mixing matrices. The residual dependence on the ratio of the two vacuum expectation values \tanb of the Higgs fields is small when \tanb is relatively large; here, \tanb is fixed to 10. In this model, $\mu$ is scanned from 100 to 240\GeV and $M_1$ from 0.3 to 1.2\TeV with the electroweakino masses and branching fractions calculated from the prescribed pMSSM parameters. The choice of parameters in the scan translates to values for the \PSGczDt-\PSGczDo mass difference that range from 4 to 28\GeV. In the scan, trilinear couplings are removed and the gluino mass parameter $M_3$ is assumed to be sufficiently high to decouple from the observable phenomena. The unification-inspired relation $M_1 = 0.5 M_2$ is further assumed, in order to reduce the parameter space to a two-dimensional scan. Cross sections are independently calculated for each model point in the pMSSM space using the \textsc{prospino2} computational package~\cite{Beenakker:1996ed}. The SUSY parameters are used to calculate the physical mass spectra and decay rates via additional computational tools~\cite{Djouadi:2002ze,Muhlleitner:2003vg,Djouadi:1997yw,Djouadi:2006bz,Skands:2003cj}.

Electroweakino decays are implemented using \PYTHIA, and are reweighted to incorporate further modeling improvements with respect to Ref.~\cite{sos2016}: the dilepton mass spectrum is reproduced from the matrix element and depends on the sign of the product of the two smallest (in magnitude) eigenvalues of the diagonalized neutralino mass matrix~\cite{DeSanctis:2007yoa}, denoted as \mczt, \mczo. While this product is always negative for a higgsino LSP, both cases are possible in the simplified wino-bino model and each is separately considered, since it leads to slightly different invariant mass distributions for the final state leptons. The branching fractions of highly virtual \PW and \PZ bosons to leptons are corrected for massive decay products, as functions of the electroweakino mass splittings, using the \textsc{susy-hit}~1.5a computational package~\cite{Djouadi:2006bz}.

Finally, two simplified models are considered for top squark pair production. The first model, denoted by \Ttbffc and motivated by Ref.~\cite{Grober:2014aha}, considers the scenario of a top squark NLSP that decays via an off-shell top quark undergoing the four-body process $\sTop\to\cPqb{f}\overline{f}\PSGczDo$, where the $f$ and $\overline{f}$ fermions are decay products of the virtual \PW. In the second model, denoted by \textsc{T2bW}, each top squark is taken to decay to a bottom quark and a chargino $\sTop\to\PQb\PSGcpmDo$, with each $\PSGcpmDo$ subsequently decaying to the LSP with a 100\% branching fraction $\PSGcpmDo\to\PW\PSGczDo$, in which the \PW boson is always off-shell. This decay dominates in naturalness-inspired models featuring light top squarks and higgsinos where only the chargino-mediated decay is not suppressed by an off-shell top quark. In this simplified model, $m_{\PSGcpmDo}=\frac{1}{2}(m_{\sTop}+m_{\PSGczDo})$ is assumed. For each scenario, a range of top squark and LSP masses are considered with a mass difference ranging from 10 to 80\GeV.

Whereas the full detector simulation is used for the wino-bino and the higgsino simplified models, a fast detector simulation \cite{cmsfastsim11,cmsfastsim14} is utilized for samples of simulated events for top squark pair production and the pMSSM higgsino model. Typically, in the investigated phase space, the fast detector simulation provides a percent-level agreement with the full detector simulation.

The trigger, lepton identification, and \PQb tagging efficiencies, as well as the distribution of pileup interactions (PU) are corrected in the simulation with scale factors measured in dedicated data samples~\cite{CMS-BTV-16-002}. Corrections for the use of the fast detector simulation are also applied to the top squark pair production and the pMSSM higgsino signal samples.

\section{Object reconstruction}\label{sec:objects}

Vertices are reconstructed from tracks according to the deterministic annealing algorithm \cite{Chabanat:865587}. The candidate vertex with the largest value of summed physics-object $\pt^2$ is taken to be the primary $\Pp\Pp$ interaction vertex (PV). The physics objects are the jets, clustered using the jet finding algorithm~\cite{Cacciari:2008gp,Cacciari:2011ma} with the tracks assigned to candidate vertices as inputs, and the associated \ptmiss, taken as the negative vector sum of the \pt of those jets. The PV must lie within 24\cm in the $z$ axis direction and 2\cm in the transverse direction from the nominal interaction point. Vertices other than the PV are associated with either PU interactions or with the decay of particles with nonnegligible lifetime (\eg, bottom quarks). The PV is then used in association with the charged-hadron subtraction algorithm \cite{PUMitigationCMS,PUMitigationJetAreas} to mitigate the effects of pileup in this analysis.

Events are reconstructed using the particle-flow (PF) algorithm~\cite{particleFlow}, which significantly improves the event description, reconstructing and identifying final-state particles by combining information gathered from the entirety of the CMS detector. The energy of photons is obtained from the ECAL measurement. The energy of electrons is determined from a combination of the electron momentum at the PV, as determined by the tracker, the energy of the corresponding ECAL cluster, and the energy sum of all bremsstrahlung photons spatially compatible with originating from the electron tracks. The muon tracks are built by the combination of measurements in the tracker and muon chambers. The momentum of a muon is obtained from the curvature of the corresponding tracks. The energy of charged hadrons is determined from a combination of their momentum measured in the tracker and the matching ECAL and HCAL energy deposits, corrected for the response function of the calorimeters to hadronic showers. Finally, the energy of neutral hadrons is obtained from the corresponding corrected ECAL and HCAL energies.

The electron momentum is estimated by combining the energy measurement in the ECAL with the momentum measurement in the tracker. It is generally better in the barrel region ($\abs{\eta} < 1.48 $) than in the endcaps ($1.48 < \abs{\eta} < 2.50$), and also depends on the bremsstrahlung energy emitted by the electron as it traverses the material in front of the ECAL~\cite{Khachatryan:2015hwa}. Muons are measured in the range $\abs{\eta} < 2.4$. Matching muons to tracks measured in the silicon tracker results in a relative muon \pt resolution of 1\% in the barrel and 3\% in the endcaps \cite{Sirunyan:2018} for \pt up to 100\GeV. 

In the current analysis, which targets the compressed region of SUSY parameter space, the final-state leptons tend to be soft: the lower \pt threshold of the electrons (muons) is set as low as 5.0 (3.5)\GeV. Only soft leptons with \pt up to 30\GeV are used in this analysis.

To identify electrons, a multivariate discriminant based on the energy distribution in the ECAL shower and track quality variables is used. Electrons must be built from tracks that have a hit at every pixel detector layer and are not associated to a conversion vertex. These requirements suppress backgrounds arising from misreconstruction or photon conversions. The identification criteria applied on the muons are based on the quality of the track in the muon system and/or the track in the tracker, and correspond to the loose and soft identification criteria of Ref.~\cite{Sirunyan:2018}.

Leptons must be isolated according to the absolute and relative isolation variables. The former counts the energy sum deposited by PF candidates in a cone of radius $\Delta R = 0.3$ around the lepton, where $\Delta R = \sqrt{\smash[b]{(\Delta\eta)^2+(\Delta\phi)^2}}$, and $\phi$ is the azimuthal angle measured in radians.
The latter isolation variable is obtained by dividing the absolute isolation by the lepton candidate \pt. The relative isolation criterion is chosen to be rather loose (less than 0.5) in order to ensure a high selection efficiency for soft leptons, while the added absolute isolation requirement is useful for candidates with higher \pt. 
To mitigate the effect of PU, only charged PF candidates with tracks associated to the PV are considered in the computation of the isolation.

Tight three-dimensional impact parameter requirements are imposed on the leptons~\cite{primaryVertex} to ensure that they are prompt, meaning that they originate from the PV. Leptons are selected only if they satisfy $\mathrm{IP}_\text{3D} < 0.01\unit{cm}$ with a significance $<2$, where $\mathrm{IP}_\text{3D}$ is the 3D distance of the lepton from the PV.

The lepton reconstruction, identification, and selection efficiencies depend on the lepton \pt and are different for electrons and muons. For electrons, the efficiency increases with \pt, starting from ${\sim}30\%$ at 5\GeV and increasing to ${\sim}70\%$ at 30\GeV. For muons, the \pt dependence is less strong and the efficiency ranges from 70 to 85\%.

Jets are reconstructed using the anti-\kt\ algorithm~\cite{Cacciari:2008gp} with a distance parameter $R = 0.4$. Tracks from charged particles not associated with the PV are removed from the clustering. Each jet is required to have $\pt \geq 25\GeV$ and to be located within the tracker acceptance ($\abs{\eta} < 2.4$). In what follows, the scalar sum of all selected jets is referred to as the transverse hadronic energy, \HT.
The energy of the jets is calibrated by correcting for PU effects, the detector response and residual differences between data and simulation~\cite{JECR}. The corrections are verified using data in dijet and $\PZ$/photon+jet events~\cite{Khachatryan:2016kdb}, exploiting the conservation of momentum in the transverse plane. 

A jet tagged as originating from the hadronization of a bottom quark is referred to as \PQb-tagged jet. The identification is achieved using the \textsc{DeepCSV} flavor tagging discriminant which combines secondary vertex and track-based information into a deep neural network~\cite{CMS-BTV-16-002}. The medium working point, which is used in this analysis, corresponds to an efficiency of about 68\% for a mistagging rate for light flavor quark and gluon jets of approximately 1\%.

The raw \ptmiss is defined as the magnitude of the negative vector \pt sum of all PF candidates reconstructed in the event \cite{Sirunyan:2019kia}, corrected by propagating the jet calibration corrections presented above. In this analysis we apply a further correction to account for the presence of muons, with the goal of matching more closely the definition used in the trigger system. We define the resulting variable as \ptmiss. 

Anomalous high-\ptmiss events can be present due to a variety of reconstruction failures, detector malfunctions, or noncollision backgrounds. Such events are rejected by event filters that are designed to identify more than 85--90\% of the spurious high-\ptmiss events with a mistagging rate of less than 0.1\%~\cite{Sirunyan:2019kia}.

\section{Event selection}\label{sec:eventselection}

The analysis requires events with a distinct signature of two or three leptons with low \pt, forming at least one OS pair and significant \ptmiss induced by an ISR jet.

The search regions (SRs) are defined in bins of raw \ptmiss and \ptmiss (for simplicity referred to as MET bins), with boundaries selected such that in each bin high efficiency and stable online selection are ensured by either the \ptmiss or the dimuon+\ptmiss trigger path. The MET binning of the analysis is presented in Table ~\ref{tab:MET_bins}. 
Four MET bins are defined for the SRs that target signal events with electroweakinos (Ewk) that decay into final states with 2 leptons (2\Pell-Ewk). Only two MET bins are considered for signal events with three leptons in the final state (3\Pell-Ewk). For signal events with top squarks (2\Pell-Stop) the upper boundaries of the MET bins are higher by 50\GeV, to increase the sensitivity of the search.  

\begin{table}[!hbtp]
    \centering
    \topcaption{Definition of the MET bins of the SRs. The boundaries of raw \ptmiss and \ptmiss (in {\GeVns}) of every bin are described.}
    \label{tab:MET_bins}
    \begin{tabular}{c c c c c c }
    \hline
         Search region & \multicolumn{2}{c}{Low-MET} & Med-MET & High-MET & Ultra-MET\\[\cmsTabSkip]
         & Raw \ptmiss & \ptmiss & \ptmiss & \ptmiss & \ptmiss \\
         \hline
         $2\Pell$-Ewk & $>125$ & $(125,200]$ & $(200,240]$ & $(240,290]$ & $>290$ \\
         $2\Pell$-Stop & $>125$ & $(125,200]$ & $(200,290]$ & $(290,340]$ & $>340$ \\
         $3\Pell$-Ewk & $>125$ & $(125,200]$ & \multicolumn{3}{c}{$>200$} \\
         \hline
    \end{tabular}
\end{table} 
 
Each MET bin is further categorized by a specific discriminating variable. In the 2\Pell-Ewk SRs, the dilepton invariant mass \mll of a pair of OS same flavor (SF) leptons, (\msfosll), is used as it has an endpoint at the mass difference of \PSGczDt-\PSGczDo. In the 3\Pell regions, the minimum of the invariant masses, (\mmnsfosll), is used as the binning variable, since we expect our signal to have small mass differences.

The \mll binning used in the low-MET bin is [4, 10, 20, 30, 50]\GeV, where the lowest boundary at 4\GeV is set due to the dimuon+\ptmiss trigger requirements described in Section~\ref{sec:sample}. Because of the requirements of this trigger, only $\PGm\PGm$ pairs are used in the low-MET bin. In the medium-, high- and ultra-MET bins, where the pure \ptmiss trigger is used, the lowest \mll boundary is relaxed to 1\GeV and the binning is [1, 4, 10, 20, 30, 50]\GeV. Dielectron pairs are also accepted in these bins.

In the 2\Pell-Stop SR, the leptons are not bound to have same flavor (except for the low-MET bin, where again only $\PGm\PGm$ pairs are allowed due to the trigger requirements mentioned above) and their \mll spectrum has no sharp endpoint. The binning in this case is applied on the \pt of the leading lepton, with boundaries [3.5, 8, 12, 16, 20, 25, 30]\GeV. 

Table~\ref{tab:CUTS} summarizes the event selection criteria applied in the three SR groups, i.e. 2\Pell-Ewk, 2\Pell-Stop and 3\Pell-Ewk, respectively.
The selection requirements based on lepton quantities are shown in the upper part of Table \ref{tab:CUTS}. Further requirements on the topology of the event are applied as listed in the lower part of Table \ref{tab:CUTS}. The motivations for some of the requirements are presented below:

\begin{itemize}
    \item An $\PGU$-meson veto and a $\PJGy$-meson veto are applied by rejecting events with \mll in the [9, 10.5] and [3, 3.2]\GeV ranges respectively.
    \item  $\HT > 100\GeV$ suppresses background events with low hadronic activity.
    \item  $2/3<(\ptmiss/\HT)<1.4$ in the dilepton final state selection is found to suppress effectively QCD multijets events, while retaining signal events boosted by ISR.
    \item The requirement of ``tight lepton veto" \cite{JetID} identification criteria for the leading jet, which removes jets from calorimetric noise as well as jets from misreconstructed leptons, in combination with the sizable \HT required, can only be realized by a jet coming from the initial state (ISR jet). This is because there is no photon or gluon final state radiation (FSR) from the LSP, which is neutral and not strongly interacting. Moreover, since the emitted SM particles and their subsequent decay products are off-shell and soft, due to the small $\Delta M$ of the signal, potential FSR from those particles or the jets from the hadronic decay of the \PW are going to have low \pt, which is much lower than the \HT required for the event selection. Finally, events with high multiplicity of soft jets that could potentially add up to 100\GeV of \HT are very unlikely. 
    \item Leptonic \PW boson decays in \ttbar events can yield two prompt leptons and $\PQb$-tagged jets in the final state. The \ttbar background is suppressed by vetoing events with $\PQb$-tagged jets with \pt above 25\GeV. Events containing softer jets from the fragmentation and hadronization of a bottom quark are still retained, e.g., in the case of the top squark decay in \Ttbffc and \textsc{T2bW} models.
    \item The contamination from Drell--Yan (DY) events can be reduced significantly by applying a selection on the approximate invariant mass of the \PZ boson. Lorentz-boosted \PZ bosons that decay into two \PGt leptons, which further decay into two leptons and four neutrinos, can satisfy the event selection requirements. In such decays, the momentum direction of the final leptons is close to the original momentum direction of the \PZ boson. The momenta of the leptons are rescaled to balance the hadronic recoil of the \PZ boson, thus yielding an estimate of the transverse momenta of the two \PGt leptons. These are then used to estimate the invariant mass of the two \PGt leptons,  $M_{\PGt\PGt}$ ~\cite{mtautau}. The range $0 \leq M_{\PGt\PGt} \leq 160\GeV$ is vetoed, since it is found to contain most of the DY events and negligible contributions from signal events. Negative values of $M_{\PGt\PGt}$ correspond to the cases when the momentum of either lepton flips direction during rescaling.
    \item The requirement on the transverse mass between each lepton and \ptmiss, \mtlm ($i=1,2$), to be less than 70\GeV, has been found to be effective in reducing the \ttbar background, the scale being set by the \PW boson mass. It is not applied in the 2\Pell-Stop SRs in order to increase the stop signal acceptance.
\end{itemize}

\begin{table}[!hbtp]
  \centering
  \topcaption{List of all criteria that events must satisfy to be selected in one of the SRs. The label ``Low-MET" refers to the low-MET bin of the analysis, while the label ``Higher-MET" refers collectively to the Med-, High- and Ultra-MET bins of the analysis.}
  \label{tab:CUTS}
  \cmsTable{
  \begin{tabular}{lcccccc}
    \hline
    \multirow{2}{*}{Variable} & \multicolumn{2}{c}{2\Pell-Ewk} & \multicolumn{2}{c}{2\Pell-Stop} & \multicolumn{2}{c}{3\Pell-Ewk}\\
      & Low-MET & Higher-MET & Low-MET & Higher-MET  & Low-MET & Higher-MET \\
     \hline
     $N_\text{lep}$ & $2$ & $2$ & $2$ & $2$ & $3$ & $3$\\
     $\pt(\Pell_{1})$ $[\GeVns]$ for e($\mu$) & $(5,30)$ & $(5(3.5),30)$ & $(5,30)$ & $(5(3.5),30)$ & $(5,30)$ & $(5(3.5),30)$ \\
     $\pt(\Pell_{2})$ $[\GeVns]$ for e($\mu$) & $(5,30)$ & $(5(3.5),30)$ & $(5,30)$ & $(5(3.5),30)$ & $(5,30)$ & $(5(3.5),30)$ \\
     $\pt(\Pell_{3})$ $[\GeVns]$ for e($\mu$) & \NA & \NA & \NA & \NA & $(5,30)$ & $(5(3.5),30)$ \\
     1 OS pair & \checkmark & \checkmark & \checkmark & \checkmark & \checkmark & \checkmark \\
     1 OSSF pair & \checkmark &  \checkmark & \checkmark & \NA & \checkmark & \checkmark \\
     $\Delta R(\Pell_i\Pell_j)$ ($i,j=1,2,3$, $i \neq j$) & \NA & $>0.3$ & \NA & $>0.3$ & \NA & $>0.3$ \\
     \msfosll (\mmnsfosll in 3\Pell) $[\GeVns]$ & $(4,50)$ & $(1,50)$ & $(4,50)$ & $(1,50)$ & $(4,50)$ & $(1,50)$ \\
     $M_\text{SFAS}^\text{max}(\Pell\Pell)$ (AS=any sign) $[\GeVns]$ & \NA & \NA & \NA & \NA & $<60$ & \NA \\
     \msfosll (\mmnsfosll in 3\Pell) $[\GeVns]$ & \multicolumn{6}{c}{veto $(3,3.2)$ and $(9,10.5)$} \\
     $\pt(\Pell\Pell)$ $[\GeVns]$ & \multicolumn{2}{c}{$>3$} & \multicolumn{2}{c}{$>3$} & \multicolumn{2}{c}{\NA} \\[\cmsTabSkip]
     Leading jet ``Tight lepton veto" & \multicolumn{2}{c}{\checkmark} & \multicolumn{2}{c}{\checkmark}& \multicolumn{2}{c}{\NA}\\
     \mtlm $[\GeVns]$ ($i=1,2$) &\multicolumn{2}{c}{$<70$}  &\multicolumn{2}{c}{\NA} & \multicolumn{2}{c}{\NA} \\
     $\HT$ $[\GeVns]$ & \multicolumn{6}{c}{$>100$} \\
     \ptmiss/\HT  & \multicolumn{2}{c}{$(2/3,1.4)$} & \multicolumn{2}{c}{$(2/3,1.4)$} & \multicolumn{2}{c}{\NA} \\
     N$_{b}(\pt>25 \GeV$) & \multicolumn{6}{c}{$=0$} \\
     $M_{\PGt\PGt}$ $[\GeVns]$ & \multicolumn{2}{c}{veto $(0,160)$} & \multicolumn{2}{c}{veto $(0,160)$} & \multicolumn{2}{c}{\NA} \\  
     \hline
     \end{tabular}
 }
\end{table}

\section{Background estimation}\label{sec:background}

The residual SM background present in the dilepton and trilepton SRs can be classified into four main categories. Two major backgrounds with prompt leptons arise from DY and \ttbar production. A third background arises from diboson production: $\PW\PW$ production for dileptons and $\PW\PZ$ production for trileptons. A fourth background arises from nonprompt or misidentified leptons, mainly from $\PW$+jets events in the dilepton search and mainly from \ttbar events in the trilepton search. Finally, rare SM processes lead to minor contributions in all SRs.

For each of the dominant prompt lepton SM backgrounds, a control region (CR) orthogonal to the SRs and enriched in the associated background process is defined. Each CR is split into two MET bins according to raw \ptmiss and \ptmiss to match the event categorization employed in the search regions:

\begin{itemize}
  \item Low-MET: Raw $\ptmiss > 125\GeV$ and $125 < \ptmiss < 200\GeV$;
  \item High-MET: $\ptmiss > 200\GeV$.
\end{itemize}

In particular, to verify and constrain the modeling of the dominant prompt-lepton backgrounds, two CRs with negligible signal contribution and with very high purity in the DY and \ttbar dilepton processes, referred to as the DY and \ttbar CRs, are used. Correspondingly, a CR designed to be enriched in trilepton $\PW\PZ$ with nonnegligible signal contamination is referred to as the $\PW\PZ$-enriched region. Another CR, with moderate purity targeting dileptons from diboson processes, VV, is introduced and is referred to as the validation region (VR). The \mll distributions from the DY CR, \ttbar CR, and $\PW\PZ$-enriched regions are included in the signal extraction procedure, which is based on a maximum likelihood fit to the data, while the VV VR is only used to assess an estimate of the corresponding background normalization uncertainty.

An additional dilepton CR comprising events with same sign (SS) leptons is used to constrain the background from nonprompt or misidentified leptons. This is described in detail in Section~\ref{sub:backgrounds:nonpromptbkg}. The SS CR is defined only for $\ptmiss >200\GeV$ (high-MET bin) and cannot be extended to lower \ptmiss, due to the opposite-sign requirement of the dimuon+\ptmiss trigger. The \mll distributions of this CR are included in the maximum likelihood fit to the data as well.

Table~\ref{tab:CR_changes} presents the selection criteria for the various background CRs that are modified with respect to those of the SR, described in Table~\ref{tab:CUTS}.

For each of the DY, \ttbar, and $\PW\PZ$ processes, and for each of the two MET bins, an unconstrained scale factor is included as a nuisance parameter in the maximum likelihood fit, to correct the normalization of the simulation yields of each process to match the data. The uncertainties on the predicted yields include the statistical and systematic components, as described in Section~\ref{sec:systematics}.

\begin{table}[!hbtp]
    \topcaption{Summary of changes in the selection criteria with respect to the SR for all the background control and validation regions.}
    \label{tab:CR_changes}
    \centering
    \begin{tabular}{cc}
        \hline
         Region & Modified selection criteria \\
         \hline
         \multirow{2}{*}{DY(2\Pell) CR}& $0 < M_{\PGt\PGt} <160 \GeV$\\
         & No upper requirement on the lepton \pt\\
         & \\
         \multirow{3}{*}{\ttbar\!(2\Pell) CR} & At least one $\PQb$-tagged jet with $\pt > 25 \GeV$\\
         & No requirement on \mtlm (instead of $<70 \GeV$)\\
         & No upper requirement on the lepton \pt\\
         & \\
         \multirow{2}{*}{VV(2\Pell) VR} & $\mtlm > 90\GeV$ (instead of $<70\GeV$)\\
         &  No upper requirement on the lepton \pt \\
         & \\
         \multirow{6}{*}{\shortstack{$\PW\PZ$(3\Pell)\\enriched region}} & No \mmnsfosll upper requirement at $50 \GeV$ \\
         & No $M_\text{SFAS}^\text{max}(\Pell\Pell)$ requirement\\
         & $\pt(\Pell_{1}) >30 \GeV$\\
         & $\pt(\Pell_{2}) >10 \GeV$ ($>15 \GeV$ if \Pell is electron in high-MET bin)\\ 
         & $\pt(\Pell_{3}) >10 \GeV$ ($>15 \GeV$ if \Pell is electron in high-MET bin) \\
         & At least one $\mu$ with $\pt>20 \GeV$ \\
         & \\
         \multirow{2}{*}{SS(2\Pell) CR} & Same-sign requirement on lepton electric charge\\
         & No requirement on \mtlm  \\
         \hline
    \end{tabular}
\end{table}

\subsection{DY control region}
\begin{figure}[!htb]
    \centering
    \includegraphics[width=0.49\textwidth]{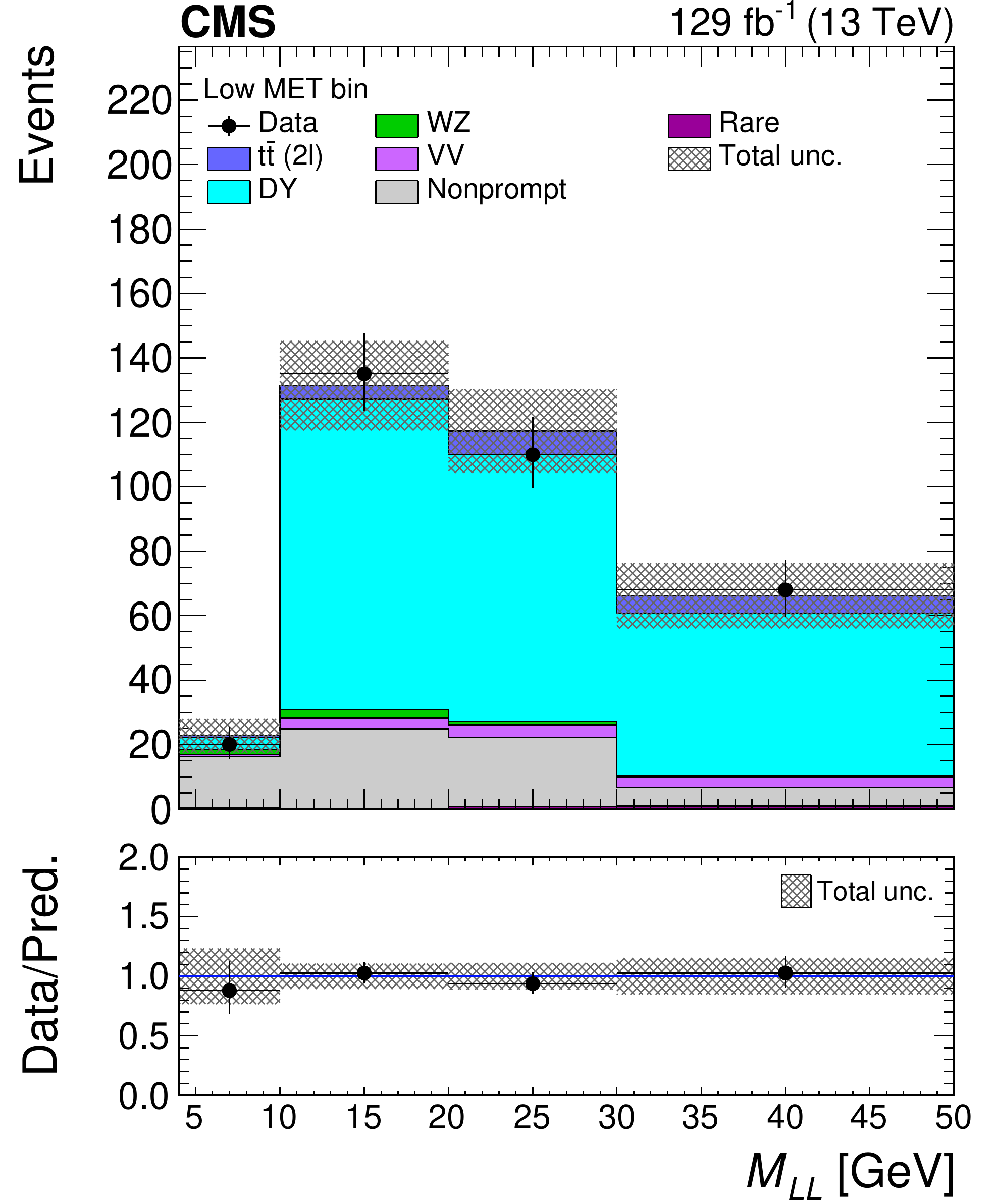}
    \includegraphics[width=0.49\textwidth]{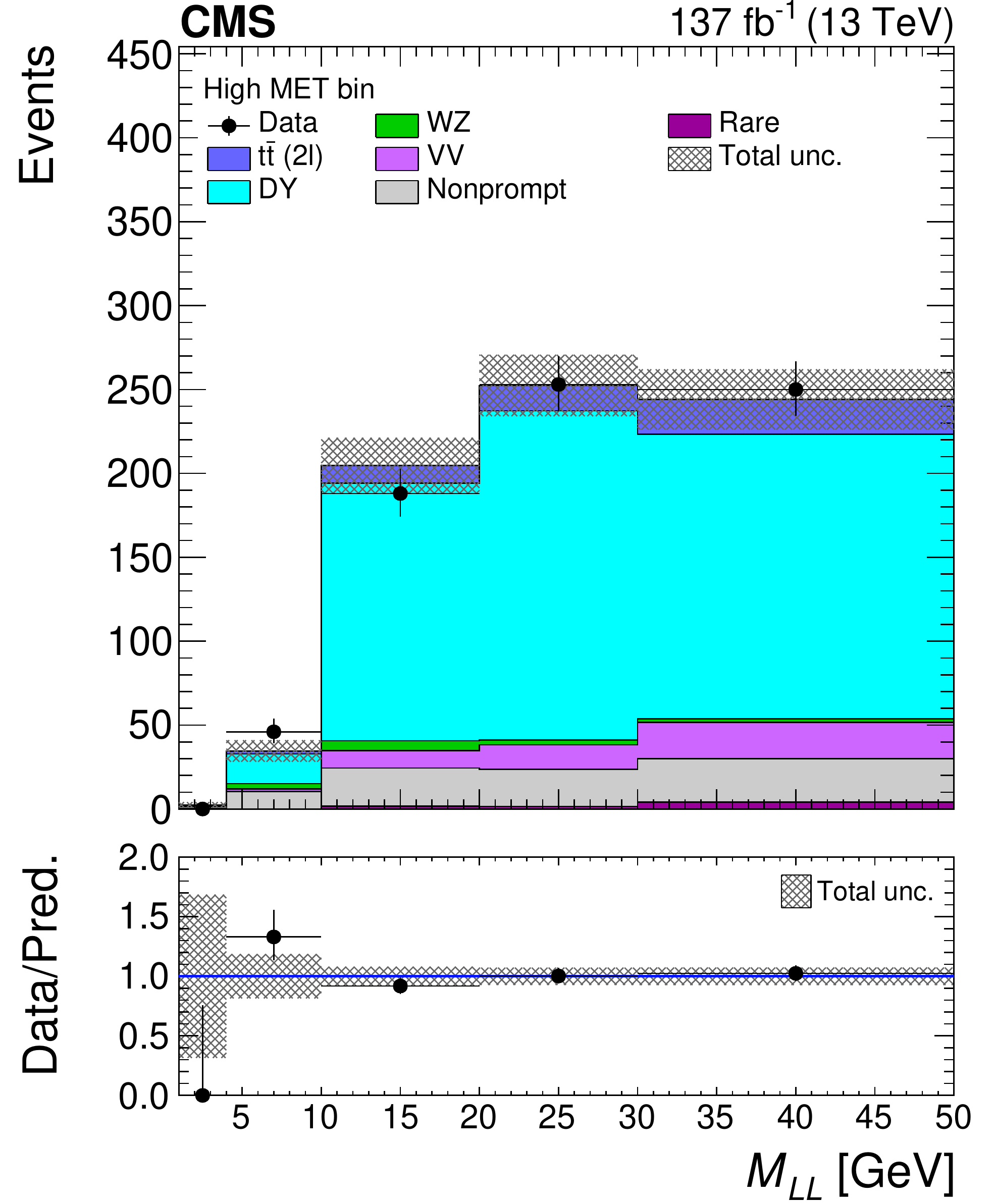}
    \includegraphics[width=0.49\textwidth]{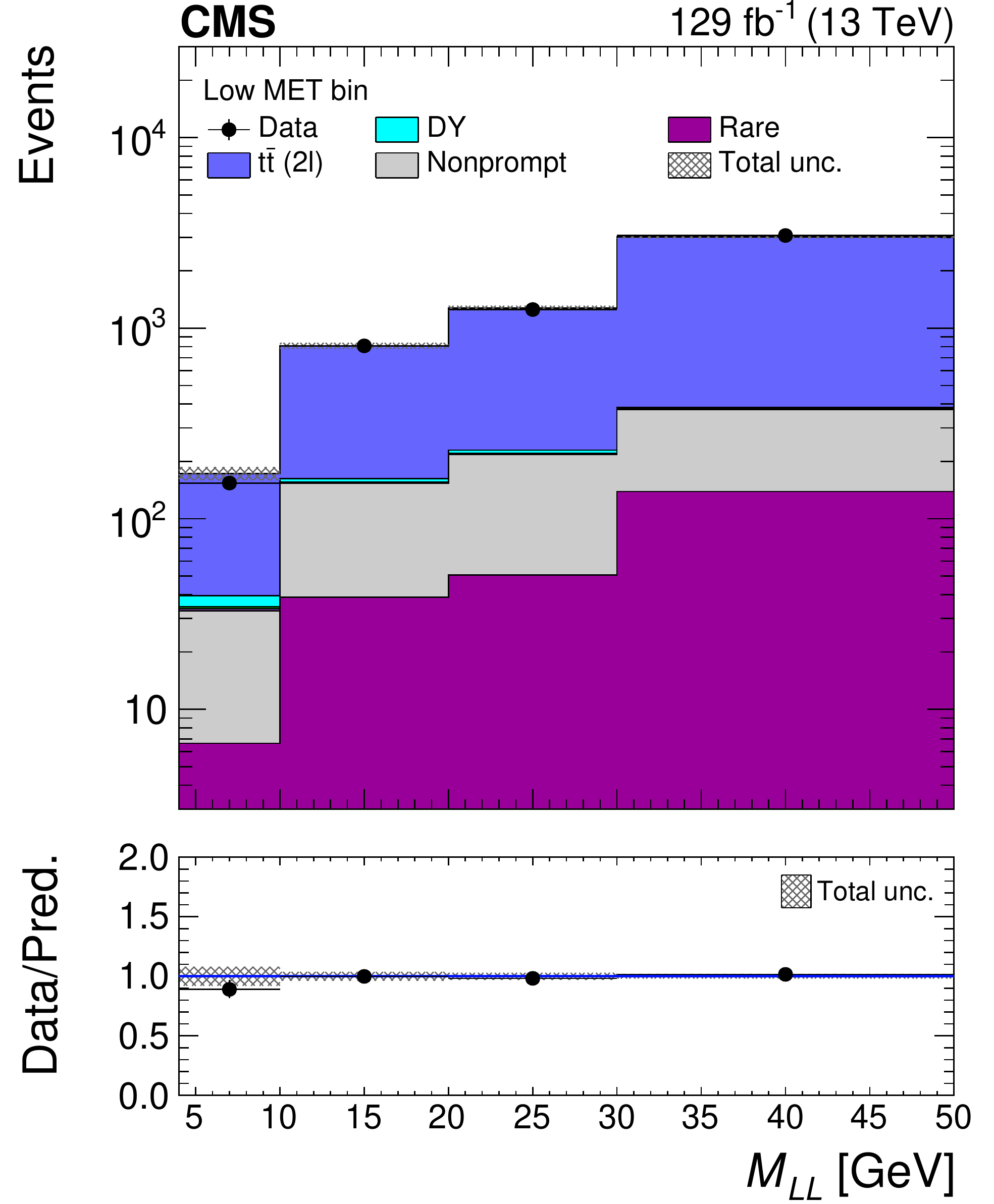}
    \includegraphics[width=0.49\textwidth]{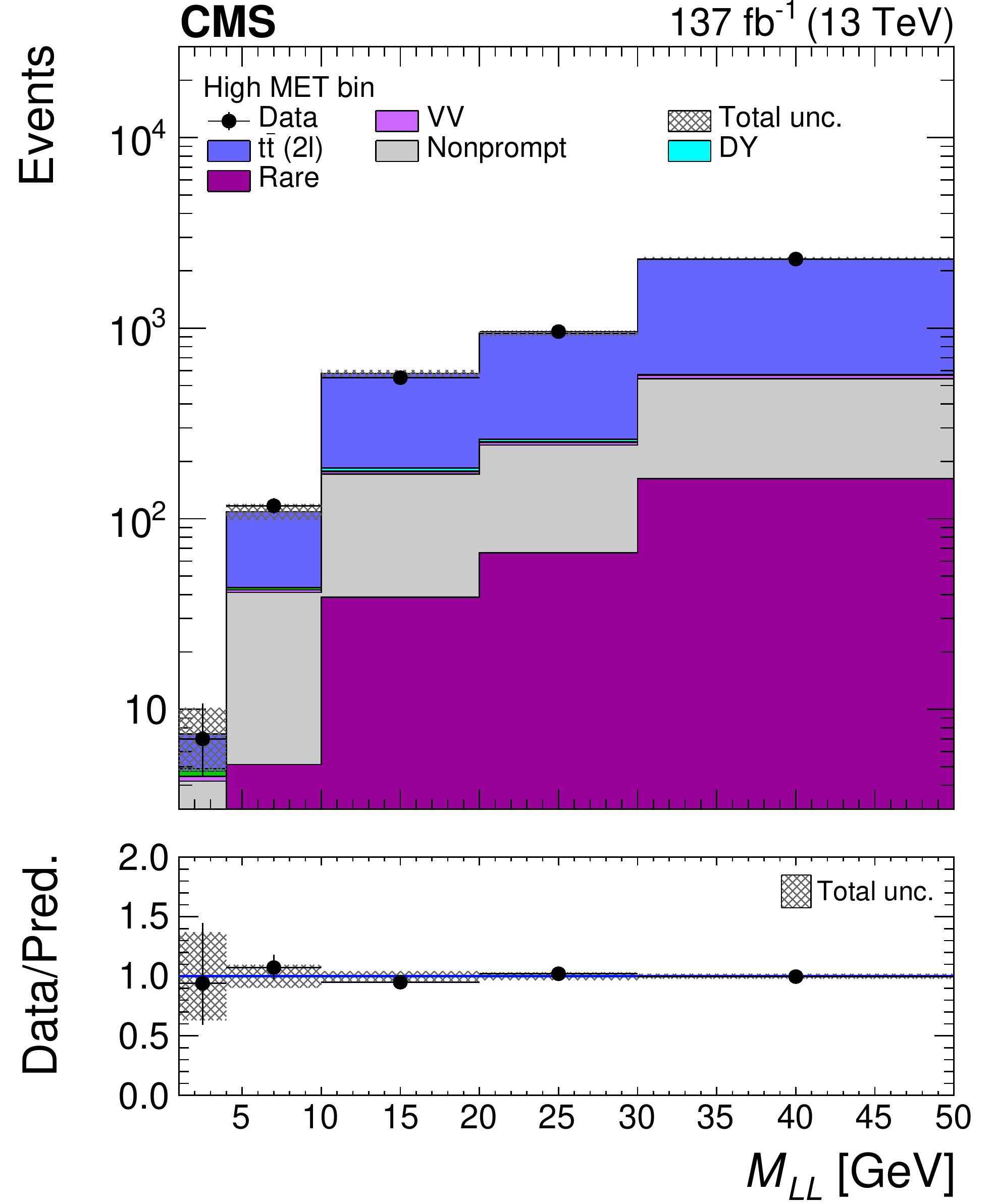}
    \caption{The post-fit distribution of the \mll variable is shown for the low- (left) and high- (right) MET bins for the DY (upper) and \ttbar (lower) CRs. Uncertainties include both the statistical and systematic components.}
    \label{fig:CRDYTT}
\end{figure}

As explained in Section~\ref{sec:eventselection}, the reconstructed mass $M_{\PGt\PGt}$ of the $\PGt$ pair is used to suppress DY events. The dedicated DY CR is defined by inverting the $M_{\PGt\PGt}$ requirement, selecting events in the range of 0--160\GeV. Additionally, the 30\GeV upper bound on the lepton \pt is removed. The post-fit distributions of the \mll variable in the DY CRs are shown in Fig.~\ref{fig:CRDYTT}. Typical scale factors of 1.2--1.7 for each data set are found between the pre- and post-fit normalizations of the DY background in the low-MET bin, and of 1.2--1.4 in the high-MET bin of the DY CR. The slightly higher deviations from unity with respect to other CRs arise because these factors correct for events that contain large amounts of instrumental \ptmiss, which is typically mismodeled in the simulation.

\subsection{\texorpdfstring{\ttbar}{tt} control region}
Even though the \ttbar background component is significantly suppressed in the search regions by vetoing events containing at least one $\PQb$-tagged jet, it is the dominant prompt background in the dilepton SRs. For the definition of the dedicated \ttbar CR at least one $\PQb$-tagged jet with $\pt > 25\GeV$ is required, ensuring orthogonality to the SR selection. In addition, the upper bound on the $\mT(\Pell,\ptmiss)$ for each lepton is removed, and no upper bound on the lepton \pt is applied. The post-fit distributions of the \mll variable are shown in Fig.~\ref{fig:CRDYTT}. Typical scale factors of 0.9--1.1 are found for each data set between the pre- and post-fit normalization of the \ttbar background in the low-MET bin, and of 0.8--1.2 in the high-MET bin.

\subsection{\texorpdfstring{$\PW\PZ$-enriched region}{WZ-enriched region}}
The decay of bosons in $\PW\PZ$ events to three leptons is the dominant prompt background process for the $3\Pell$ SRs (the $\PW\PZ$ label shown in the plots corresponds to \PW and \PZ bosons both decaying leptonically while the other decay modes are included in the VV label, described in the next subsection). In order to assess the normalization of the SM $\PW\PZ$ process to fully leptonic final states, a $\PW\PZ$-enriched region is employed, split in the same two MET bins as for the other CRs. In the low-MET bin of the $\PW\PZ$-enriched region, a pure dimuon trigger is used instead of the dimuon+\ptmiss one, since the latter includes a requirement on the invariant mass of the two muons. The event selection in the $\PW\PZ$-enriched region differs from the 3\Pell SR event selection in that no upper \mmnsfosll selection or $M^{\text{max}}(\Pell\Pell)$ \PZ veto is applied. Furthermore, no upper bound on $\pt(\Pell)$ is applied, while the lower bounds on $\pt(\Pell)$ are set to $\pt(\Pell_1) >30\GeV$, $\pt(\Pell_2) >10\GeV$ and ${\pt(\Pell_3)>10 \GeV}$, requiring at least one muon with ${\pt > 20 \GeV}$ as driven by the pure dimuon trigger. The fraction of events from $\PW\PZ$ decays in this selection is $78(88)\%$ in the low(high)-MET bin.

When split into the \mll bins, the low-\mll range of the $\PW\PZ$-enriched region could have some nonnegligible signal contribution in the case of intermediate- and higher-mass splittings (30--40\GeV) with respect to the overall $\PW\PZ$ process contribution. Therefore, the part of the $\PW\PZ$-enriched region that satisfies $1 < \mll < 30\GeV$ contributes to the sensitivity of the analysis and is defined as $\PW\PZ$-like SR. The $\mll > 30\GeV$ region, which includes the majority of the $\PW\PZ$ process yields, is defined as $\PW\PZ$ CR.

\begin{figure}[!htb]
    \centering
    \includegraphics[width=0.49\textwidth]{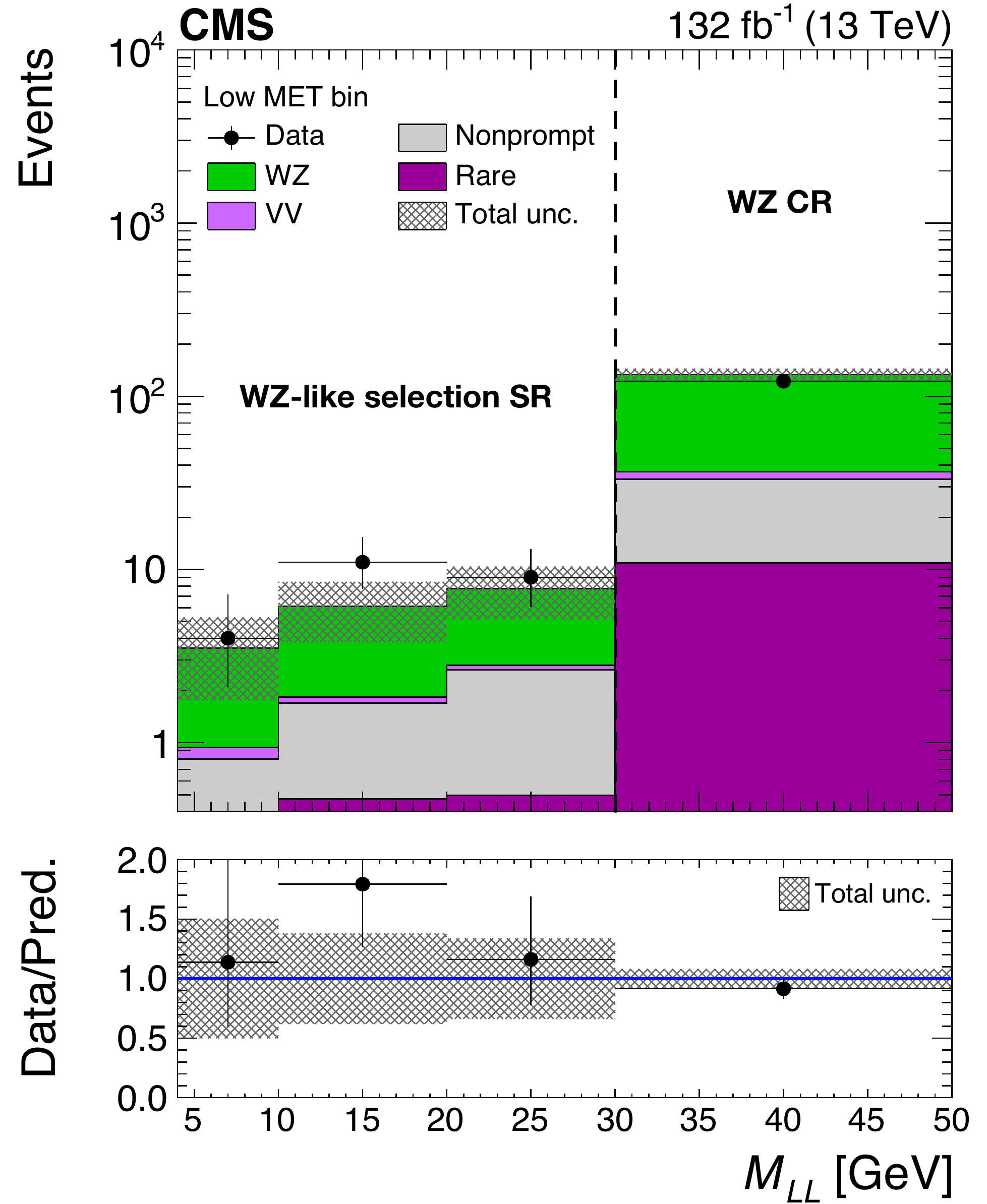}
    \includegraphics[width=0.49\textwidth]{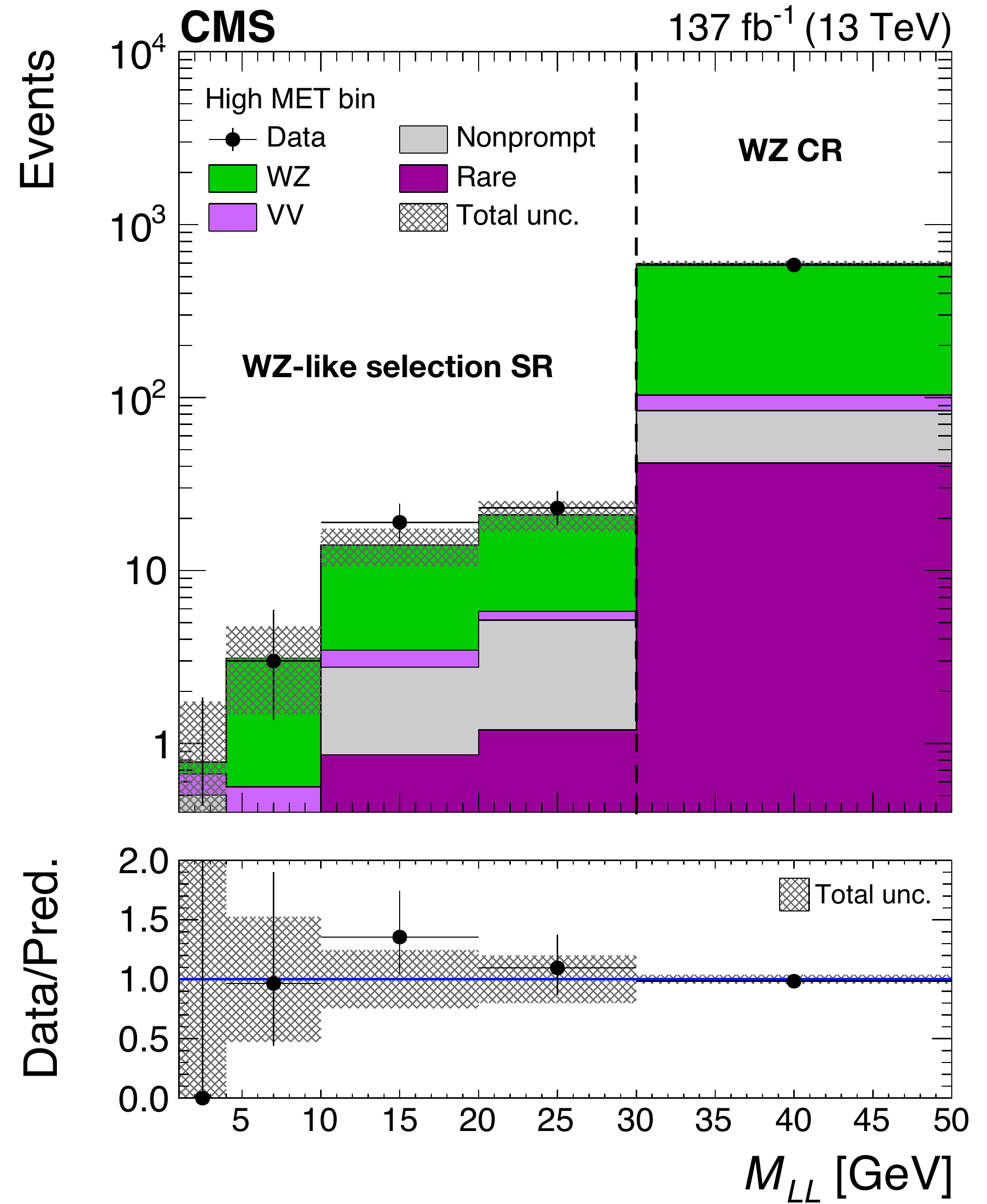}
    \caption{The post-fit distribution of the \mll variable is shown for the low- (left) and high- (right) MET bins for the $\PW\PZ$-enriched region. Uncertainties include both the statistical and systematic components.}
    \label{fig:CRWZ}
\end{figure}

The post-fit distributions of the \mll variable are shown in Fig.~\ref{fig:CRWZ}. Typical scale factors of 0.7--0.8 are found for each data set between the pre- and post-fit normalizations of the $\PW\PZ$ background in the low-MET bin, and of 0.6--0.8 in the high-MET bin. These scale factors tend to have somewhat large deviations from unity, as they account for the overestimation of the $\PW\PZ$ background normalization in the simulation.

\subsection{VV validation region}
\label{sub:backgrounds:vv}

Diboson production VV accounts for the mixture of $\PW\PW$, $\PW\PZ$ (all decay modes, except the fully leptonic one), and $\PZ\PZ$ events, where the processes are given in descending order with respect to their contribution in the dilepton search regions. The event selection of the VV VR is obtained by inverting the \mtlm requirement, removing the upper bound on the lepton \pt and requiring a high-\pt ($>30\GeV$) leading lepton to ensure orthogonality to the SRs.
This region is less pure with respect to the CRs mentioned so far; it has a purity of 18\% and 31\% for the low-MET and high-MET selection, respectively. It is thus only used to check the pre-fit data and simulation agreement, which is found to be good. This VV-like region is not included in the maximum likelihood fit.

\subsection{Rare SM processes}
The previously mentioned SM processes leading to minor contributions in the SRs are collectively referred to as rare. These comprise the production of a {\PW}, {\PZ} or Higgs boson ({\PH}) in association with top quarks, such as $\PQt\PW$, $\ttbar$V, $\ttbar\PH$, $\PQt\PZ\PW$, and triboson processes VVV.
Due to their very small contribution in the SR, no dedicated CR is designed for the estimation of these processes and their yield is taken directly from simulation.

\subsection{Nonprompt background}
\label{sub:backgrounds:nonpromptbkg}

The background from nonprompt or misidentified leptons is evaluated with the ``tight-to-loose" method \cite{Khachatryan:2017qgo}. For the calculation of the nonprompt background in the SR two additional regions need to be defined. The application region (AR) is selected by applying the SR requirements except the tight identification and isolation lepton requirement which is replaced by at least one lepton failing the tight identification and isolation criteria but passing a looser selection. Therefore, the AR is enriched in nonprompt leptons. The measurement region (MR) is a region enriched in events that contain strongly produced jets, referred to as QCD multijet events. The MR is defined by requiring one loose lepton, and a jet separated from the lepton by $\Delta R \geq 0.7$. The probability that a nonprompt lepton that passes the loose selection also satisfies the tight selection is called the fake rate and it is measured in MR data as a function of lepton \pt and $\eta$. The probability of a prompt lepton to pass the tight identification and isolation requirement is called prompt rate and it is measured in simulation. The nonprompt background estimation in the SR is performed by weighting the AR events by a transfer factor (TF), which depends on the fake rate and the prompt rate.

The MR for muons is selected by prescaled single-muon triggers with no isolation requirement, while for electrons, a mixture of prescaled jet triggers is used. The jet selection requirements applied for the MR for muons is $\pt \geq 50 \GeV$ and for the electrons are $\pt \geq 30 (40)\GeV$ in 2018 (2016 and 2017), according to the trigger requirements that evolved during data-taking. 

In the data-driven tight-to-loose method, the fake rate is measured in QCD multijet data events of the MR and applied in $\PW$+jets and \ttbar data events with loose leptons in the AR. Therefore, the jet flavor composition of the MR and the AR may differ and this can affect the prediction of the nonprompt background. The fake rate calculated in QCD multijet simulation was studied and found to be consistent across the different processes. The closure of the method is verified by applying the fake rate determined from QCD multijet data events to simulated $\PW$+jets and \ttbar events containing at least one nonprompt lepton in the AR. The resulting estimated nonprompt background in the SR is compared with the observed simulated $\PW$+jets and \ttbar events with at least one nonprompt lepton in the SR. The maximum nonclosure of the method is found to be 40\% and this value is used as the systematic uncertainty in the normalization of the nonprompt background.

The TF are constructed with the prompt rate and the fake rate and they are applied on the AR events, in which at least one lepton fails the tight identification and isolation criteria. This yields the number of events with nonprompt leptons in the SR. This method is applied in MET bins with sizable numbers of events, \eg, the low-MET bins of the 2\Pell-Ewk and Stop SRs. In MET bins with limited numbers of events, namely the medium-, high- and ultra-MET bins of the 2\Pell-Ewk and Stop SR and the low- and high-MET bins of the 3\Pell SR, the nonprompt-lepton background is estimated by applying the TF to the \mll shapes in the AR, as obtained by simulation and normalized to data. This latter approach maintains the robustness in the measurement of the misidentification probability by utilizing control samples in data and in the normalization of the misidentified-lepton background in the AR, while using the \mll shape in simulation to reduce statistical fluctuations.

To reduce the statistical uncertainty in the estimation of the nonprompt-lepton background in the 2\Pell-Ewk medium-, high- and ultra-MET SRs, the \mll shape of the nonprompt-lepton simulation in the AR is merged into one inclusive MET bin. The simulated \mll shape of the nonprompt-lepton background is indeed found to be compatible across different MET bins. A dedicated systematic uncertainty is applied to the nonprompt-lepton background in the 2\Pell-Ewk medium-, high- and ultra-MET bins, to account for the \mll shape extrapolation across the MET bins.

\begin{figure}[!htb]
    \centering
    \includegraphics[width=0.49\textwidth]{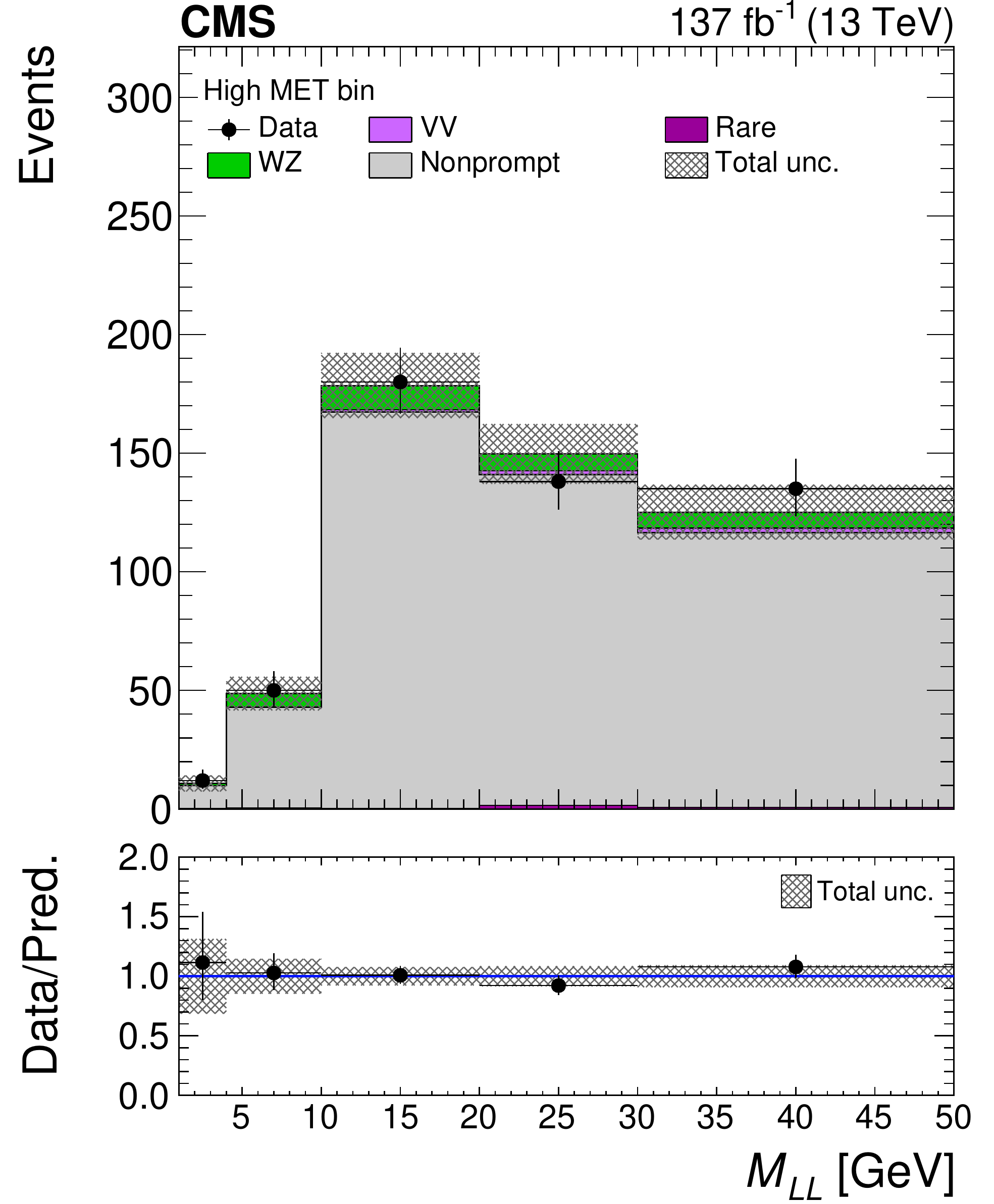}
    \caption{The post-fit distribution of the \mll variable is shown for the high-MET bin for the SS CR. Uncertainties include both the statistical and systematic components.}
    \label{fig:CRSS}
\end{figure}

The SS CR is used to further constrain the nonprompt-lepton background prediction uncertainty using data. This CR is obtained by requiring two leptons of the same sign instead of opposite sign when applying the 2\Pell-Stop SR selection in the $\ptmiss>200\GeV$ region. The requirement of two SS leptons increases significantly the probability that at least one of the two is nonprompt, thus enriching the CR in nonprompt-lepton background. The \mll distribution of the SS CR, with the nonprompt-lepton background predicted with the tight-to-loose method, is included in the maximum likelihood fit. The post-fit \mll distribution of the SS CR is shown in Fig.~\ref{fig:CRSS}. A scale factor of 1.06 is estimated between the pre- and post-fit normalizations of the tight-to-loose prediction.

\section{Systematic uncertainties}\label{sec:systematics}

Systematic uncertainties arise from experimental effects and from the modeling of the simulated processes. This section describes the sources of systematic uncertainties and quantifies their effect on the estimated background. The post-fit effect of each source of uncertainty is quoted.

All of the trigger, lepton selections, and \PQb tagging efficiency corrections that are applied to simulated samples have uncertainties related to the methods used to estimate them from data; these uncertainties affect the final predictions with values that fall in the ranges 2--9\%, 2--9\% and 1--4\%, respectively. Additional experimental uncertainties arise from the jet energy scale and resolution corrections that account for any differences between data and simulation when measuring the jet energies. These uncertainties affect all simulated backgrounds and result in 1--10\% uncertainty in the total background prediction. The uncertainties of the trigger efficiency and lepton selection are applied separately for each year, while the \PQb tagging efficiency and the jet energy corrections are correlated across years and are treated accordingly.

Dedicated weights are applied to the simulated samples to correct the distribution of the true number of PV to the one measured in data. The number of interactions per bunch crossing is estimated from the total inelastic cross section, which has been measured with an uncertainty of 4.6\% \cite{PUMeasurement}. The PU uncertainty is estimated by propagating the total inelastic cross section uncertainty to the PU weights and using their variation as a shape uncertainty for all years combined. The net effect on the final total background prediction is found to be at most 1--2\%.

The uncertainties in the luminosity measurements are incorporated in the estimates of all the prompt-lepton backgrounds and the predicted signal yields. These uncertainties are 2.5\%, 2.3\% and 2.5\% for the 2016~\cite{CMS-PAS-LUM-17-001}, 2017~\cite{CMS-PAS-LUM-17-004}, and 2018~\cite{CMS-PAS-LUM-18-002} data set, respectively, treated as uncorrelated among the different years.

As presented in Section 6, the dedicated CRs for the dominant prompt-lepton backgrounds are included in the maximum likelihood fit to the data. The normalization of each of these processes is left to float freely in the fit, independently for each year and for each MET bin. The resulting uncertainties are found to be in the ranges 15--35\% for the DY, ${\sim}15\%$ for the \ttbar, and 12--27\% for the $\PW\PZ$ backgrounds.

The VV background, described in Section~\ref{sec:background}, accounts for all the diboson processes that enter the dilepton search. Its modeling is validated in the VV VR, and its estimated contribution is assigned an uncertainty of 50\%, correlated across the three years.

To fully account for all uncertainties related to the simulation of the rare SM processes, a conservative uncertainty of 50\% has been assigned to these minor backgrounds for all years.

From the closure tests of the tight-to-loose method, which are presented in Section~\ref{sub:backgrounds:nonpromptbkg}, a pre-fit systematic uncertainty of 40\%, correlated across the three years of data-taking, is assigned on the misidentified-lepton background estimation. This background, however, is constrained significantly by including the SS CR in a single MET bin ($\ptmiss > 200\GeV$) in the maximum likelihood fit. With the inclusion of the SS CR in the fit, the post-fit uncertainty in the nonprompt-lepton background after the fit is reduced to ${\sim}5\%$.

An extra shape uncertainty is applied to the nonprompt-lepton background prediction to account for potential disagreements between data distributions and the templates from simulation used in the ARs. An additional shape uncertainty is included for the SRs for which the corresponding AR distributions are merged across MET bins to account for minor shape discrepancies in different MET bins. These shape uncertainties are applied to the nonprompt prediction independently for every year. The post-fit effect of these two separate shape uncertainties on the nonprompt background is approximately 8\% and 5\%, respectively.

Two additional uncertainties are assigned to all simulated signal samples: The uncertainty in the theoretical cross section due to the variation of the parton density functions is included in the ${\pm}1\sigma$ curves on the limit scans, and amounts to 3.5--8.5\%. The acceptance uncertainties due to the variations of the renormalization and factorization scales are added separately as nuisance parameters in the maximum likelihood fit and are of ${\sim}1\%$ each.

Potential differences between data and simulation in modeling ISR jets are also a source of systematic uncertainty. To this end, the data are compared with the simulation in a general selection that requires two isolated leptons and at least one ISR jet. The residual differences between data and simulation are used to determine \pt-dependent correction factors, independently for each data-taking year. The correction factors are applied to the simulation and their differences from unity (1--5\%) are assigned as a systematic uncertainty.

Differences in the \ptmiss reconstruction in the detailed and fast simulations used for the signal samples are taken into account as an additional systematic uncertainty for each year separately. This uncertainty varies between 1 and 10\% for the yields of the fast simulation signal samples.

During the 2016 and 2017 data-taking periods, a gradual shift was observed in the timing of the trigger information from the ECAL towards early values. This causes a sizable fraction of electromagnetic objects with $\abs{\eta} > 2.5$ to be assigned to the previous bunch crossing. To account for this issue, correction factors are applied and their uncertainties ($\sim$1\%) are propagated to the final result.

The uncertainties presented above are included as nuisance parameters in the likelihood fit to the data. To freely float in the fit, the DY, \ttbar and $\PW\PZ$ backgrounds are assigned uniformly distributed uncertainties, whereas all other fit parameters are assigned normally distributed uncertainties. The dominant uncertainties are generally the statistical ones. The prompt and nonprompt background normalization uncertainties also become important, depending on the part of the phase space that is probed.

\section{Results}\label{sec:results}

The signal and background expectations are fitted to the data using a binned maximum likelihood fit with the uncertainties incorporated as nuisance parameters, as mentioned above. The signal yields are scaled in all regions of the analysis (also in the CRs to account for possible signal contamination) by a single, unconstrained parameter of interest. The CRs for the DY, \ttbar and $\PW\PZ$ processes are added to determine the normalization of the respective processes. Similarly, the SS dilepton CR is included in the fit in order to constrain the nonprompt background. The distribution that enters the maximum likelihood fit is the \mmnsfosll for the 3\Pell CRs and the 3\Pell-Ewk SRs, and the \mll for the 2\Pell CRs and the 2\Pell-Ewk SRs. Each distribution is comprised by 4 bins in the low-MET bins and 5 bins otherwise. For the 2\Pell-Stop SRs, the \pt distribution of the leading lepton (6 bins in each MET bin) is used.
In total, the fit that targets signal models with electroweakinos is given 28 SRs (19 dilepton and 9 trilepton SRs), while for models with top squark production 24 SRs (dilepton only) are used. In both cases, the same 32 CR bins are used.

\begin{figure}[!hbtp]
    \centering
    \includegraphics[width=0.49\textwidth]{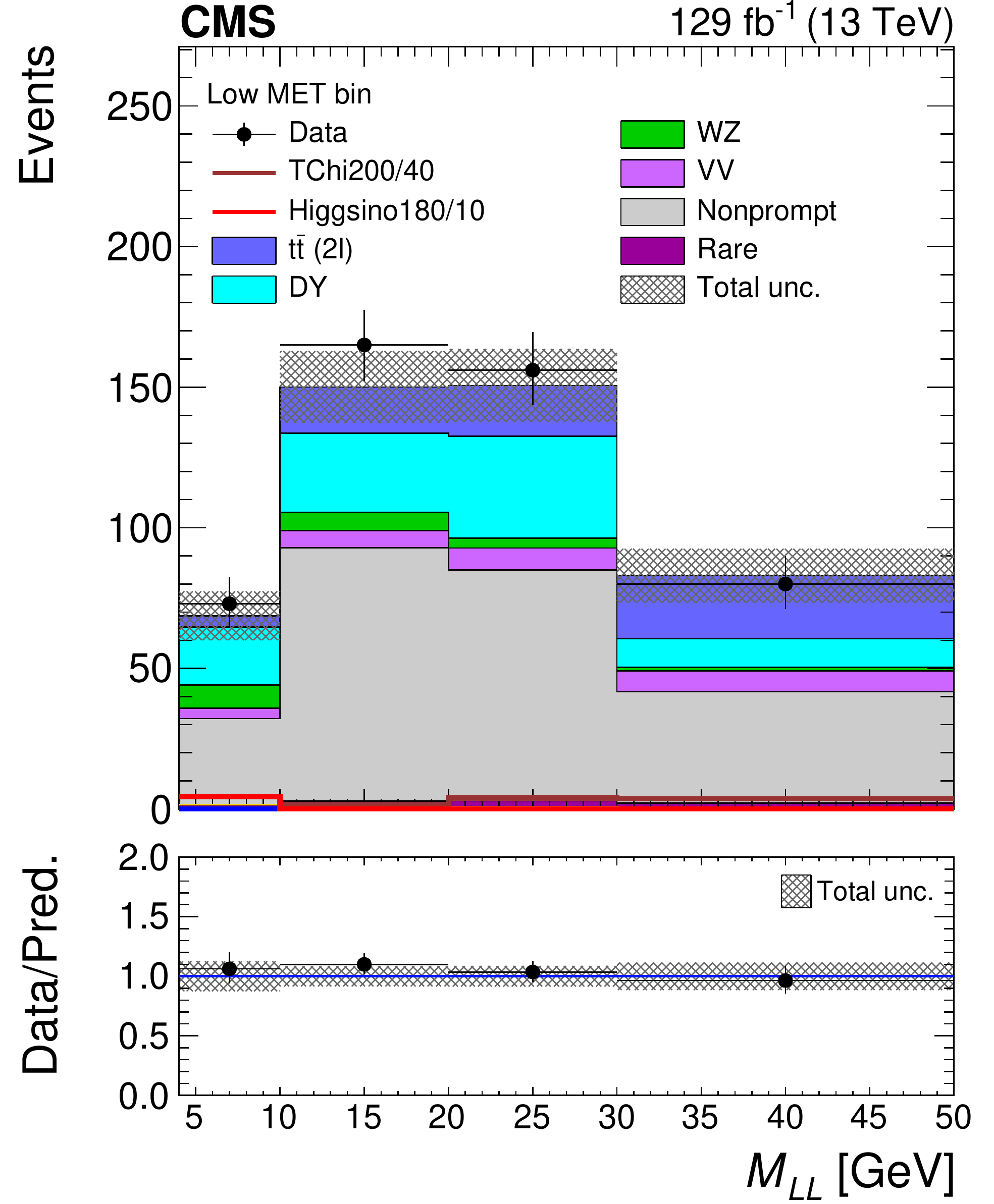}
    \includegraphics[width=0.49\textwidth]{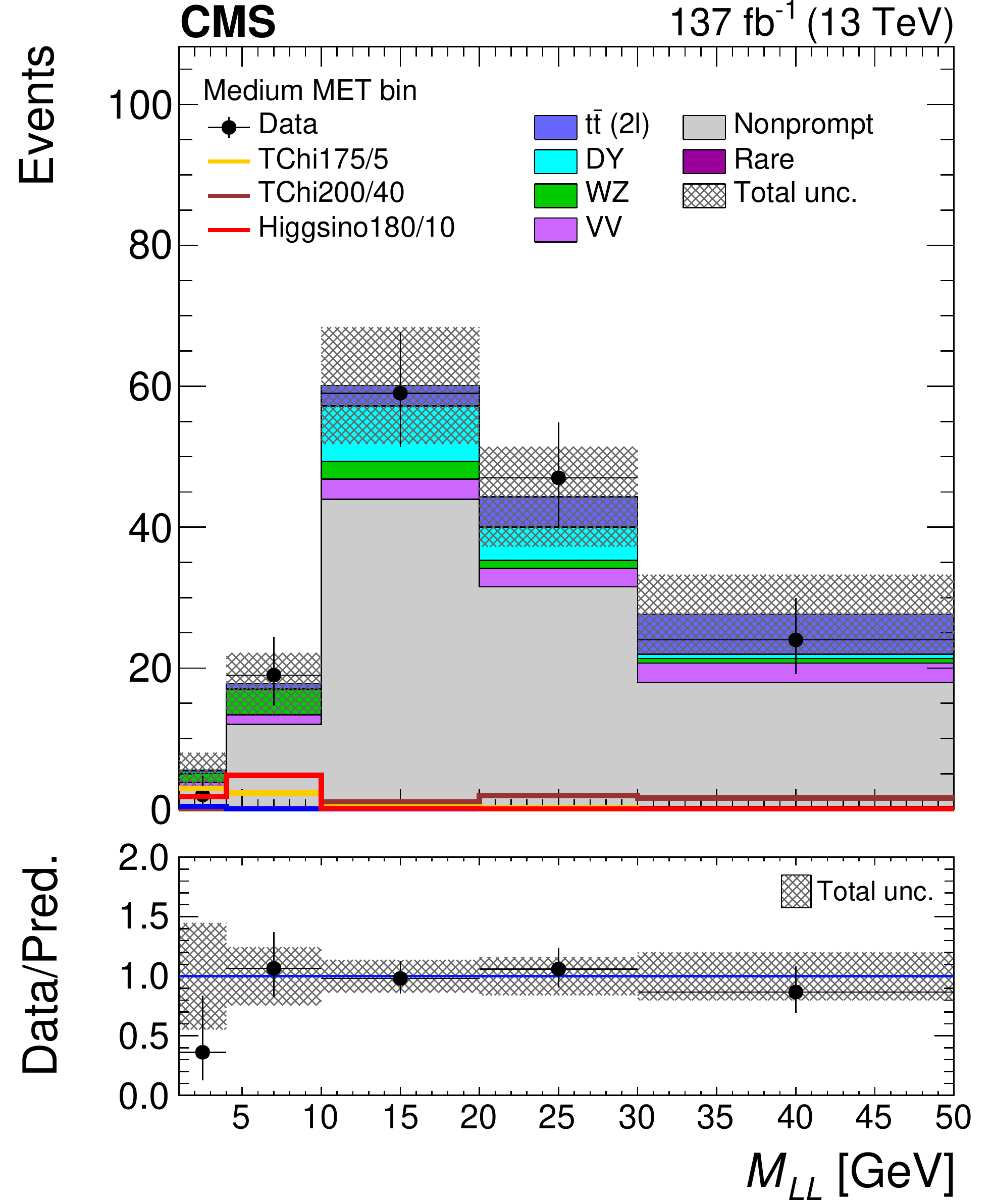}
    \includegraphics[width=0.49\textwidth]{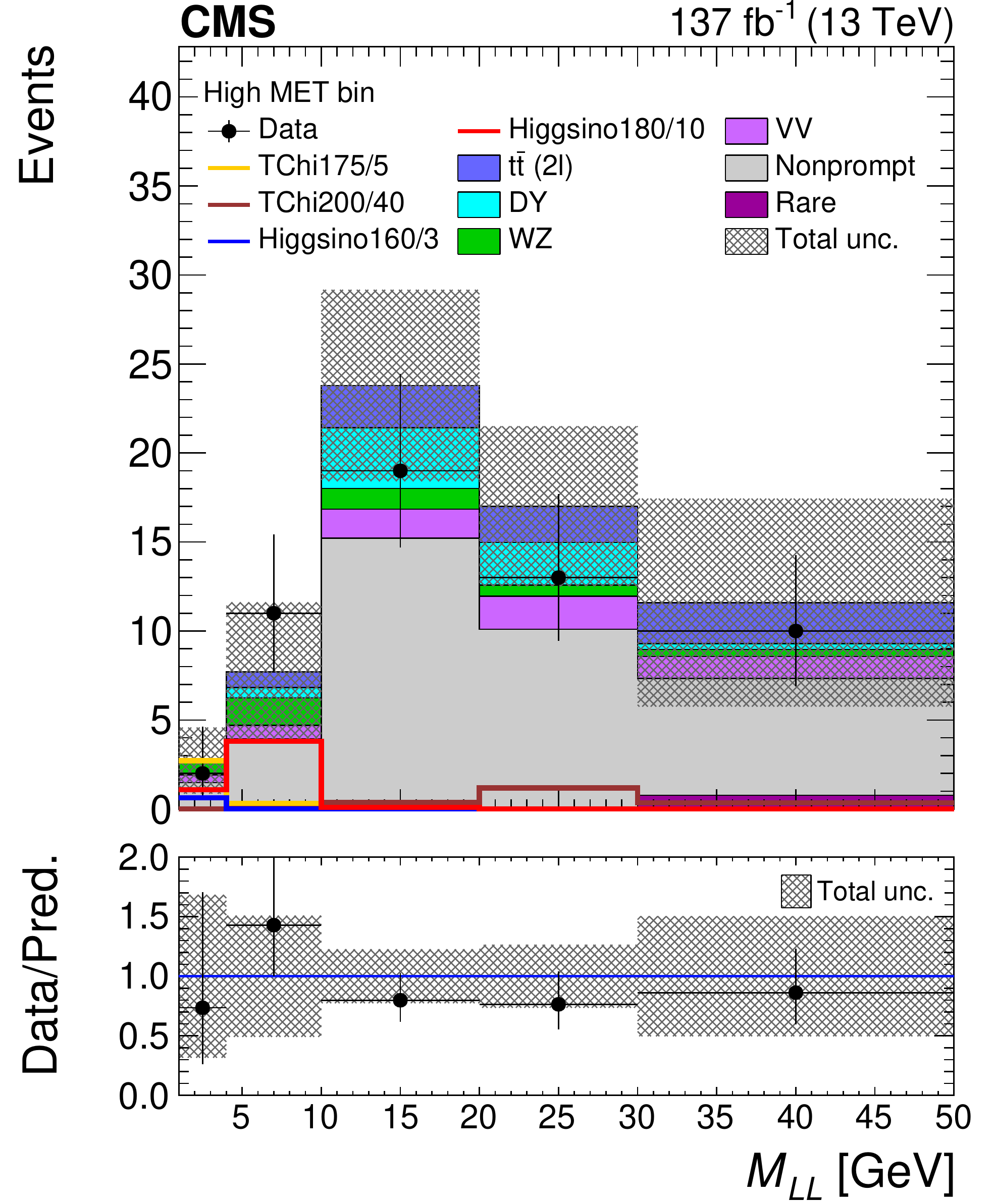}  
    \includegraphics[width=0.49\textwidth]{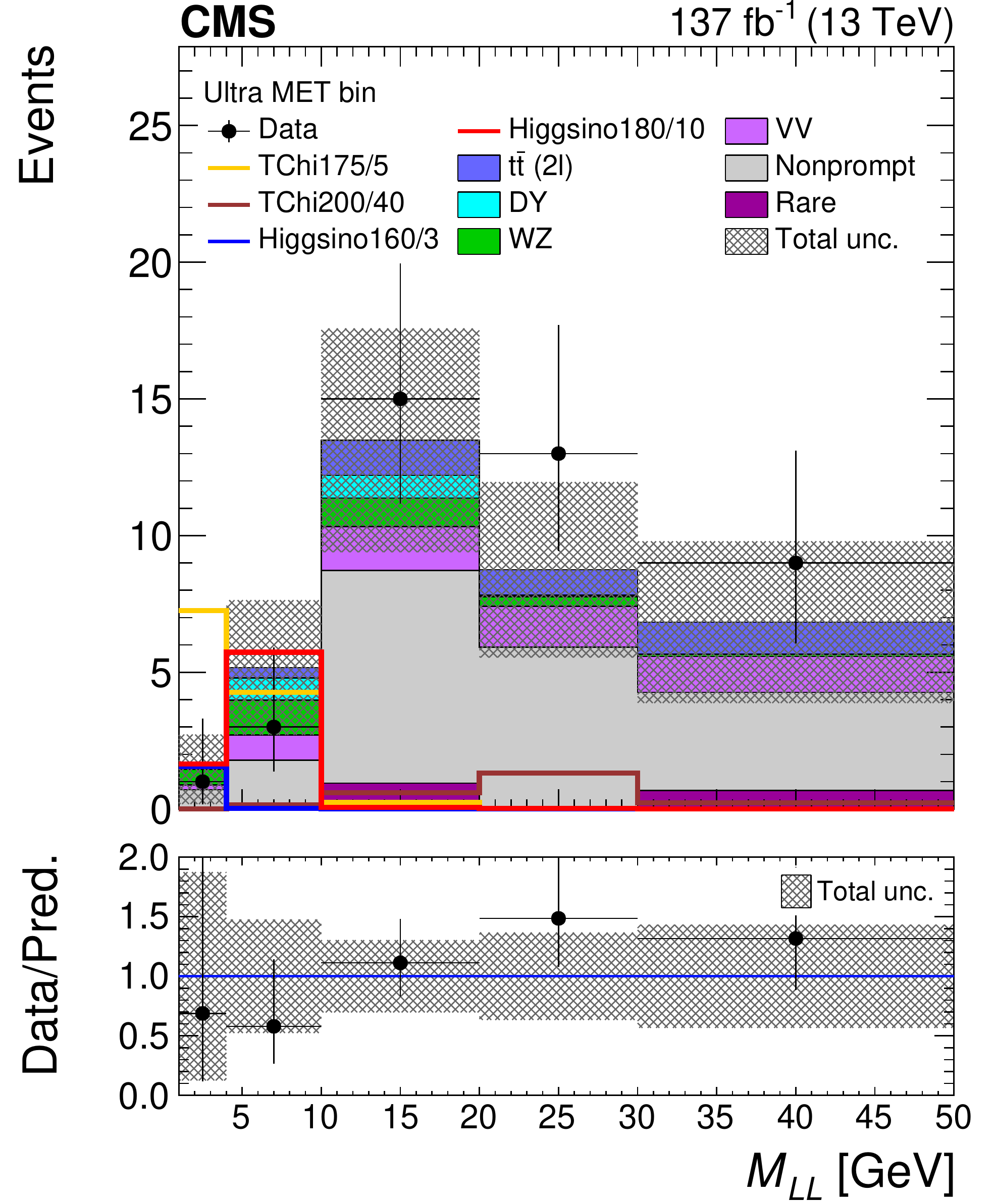} 
\caption{The 2\Pell-Ewk SR: the post-fit distribution of the \mll variable is shown for the low- (upper left), medium- (upper right), high- (lower left) and ultra- (lower right) MET bins. Uncertainties include both the statistical and systematic components. The signal distributions overlaid on the plot are from the \TChiWZ and the simplified higgsino models in the scenario where the product of $\mczo\mczt$ eigenvalues is positive and negative, respectively. The numbers after the model name in the legend indicate the mass of the NLSP and the mass splitting between the NLSP and LSP, in {\GeVns}.}
\label{fig:results:2lEwkSR}
\end{figure}

\begin{figure}[!hbtp]
  \centering
    \includegraphics[width=0.49\textwidth]{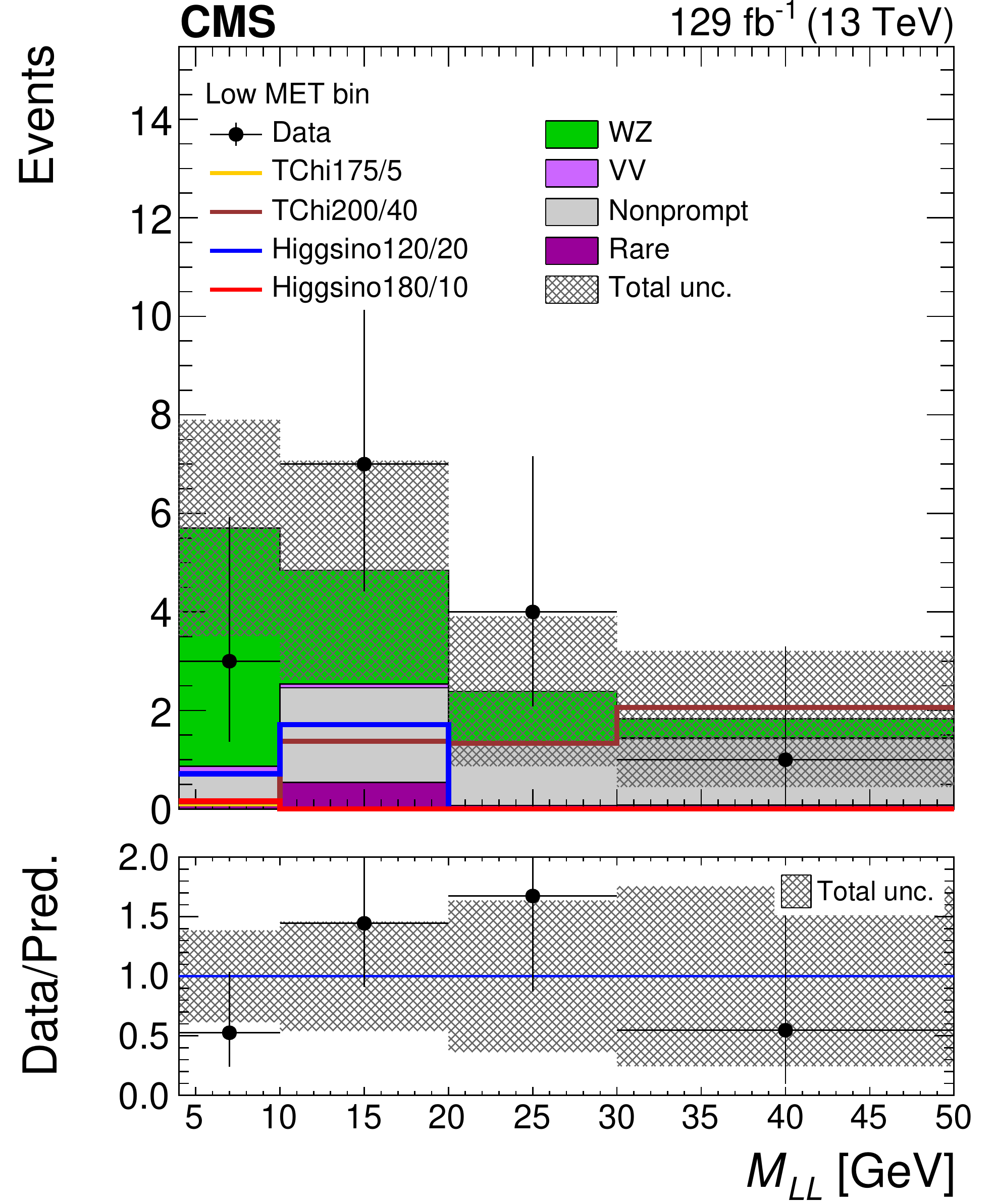}
    \includegraphics[width=0.49\textwidth]{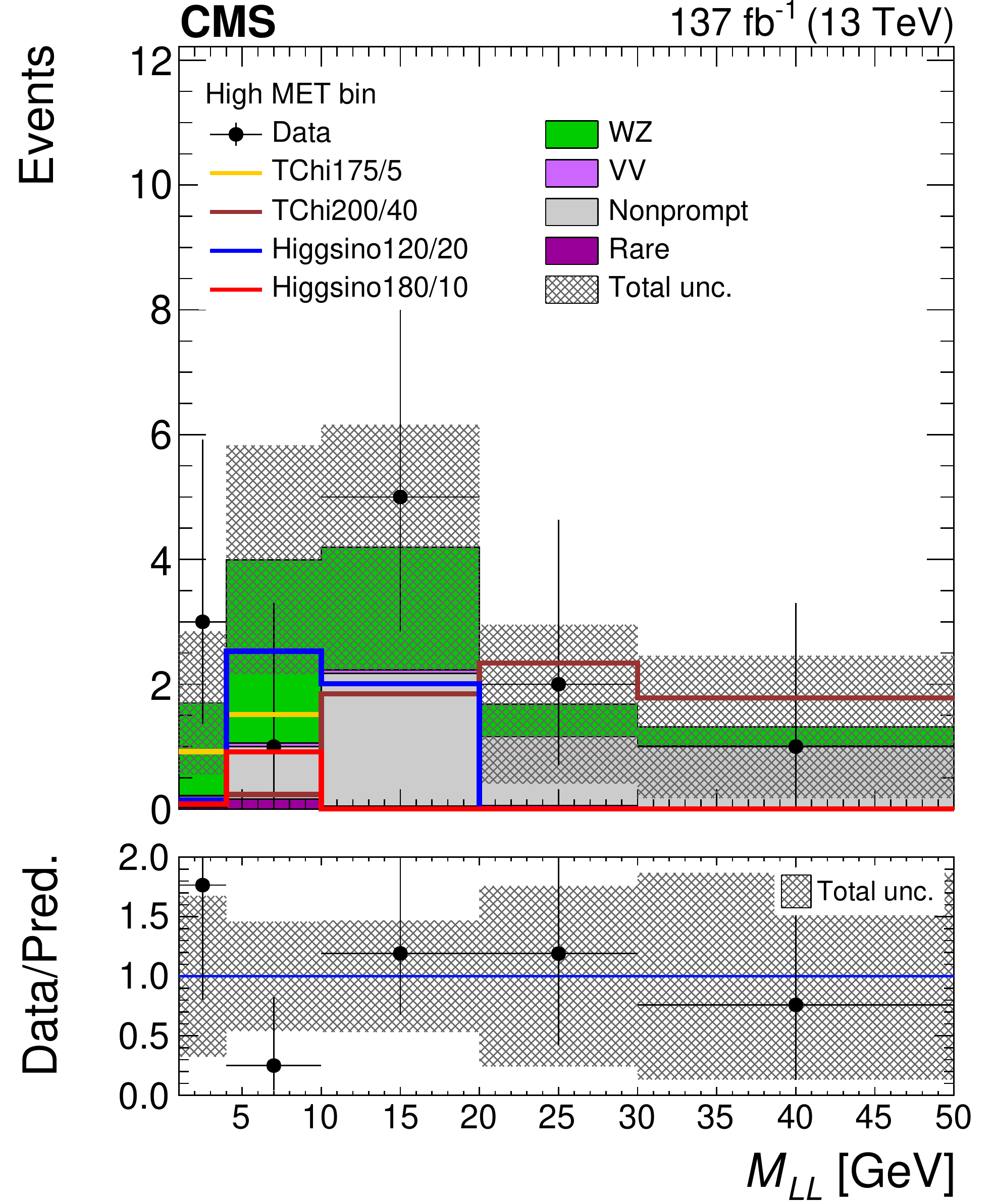}
\caption{The 3\Pell-Ewk search regions: the post-fit distribution of the \mmnsfosll variable is shown for the low- (left) and high- (right) MET bins. Uncertainties include both the statistical and systematic components. The signal distributions overlaid on the plot are from the \TChiWZ and the simplified higgsino models in the scenario where the product of $\mczo\mczt$ eigenvalues is positive and negative, respectively. The numbers after the model name in the legend indicate the mass of the NLSP and the mass splitting between the NLSP and LSP, in {\GeVns}.}
  \label{fig:results:3lSR}
\end{figure}

\begin{figure}[!hbtp]
    \centering
    \includegraphics[width=0.49\textwidth]{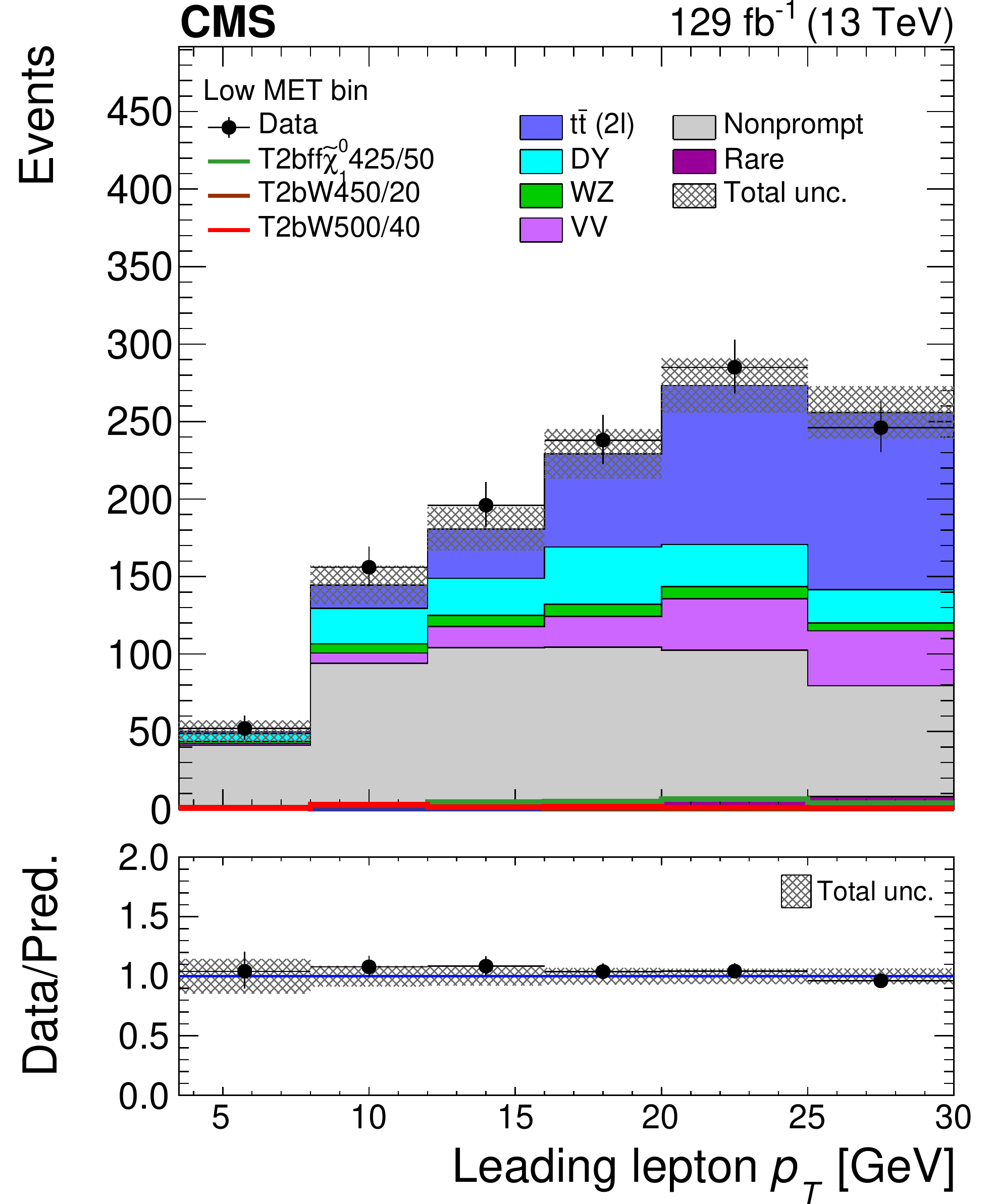}
    \includegraphics[width=0.49\textwidth]{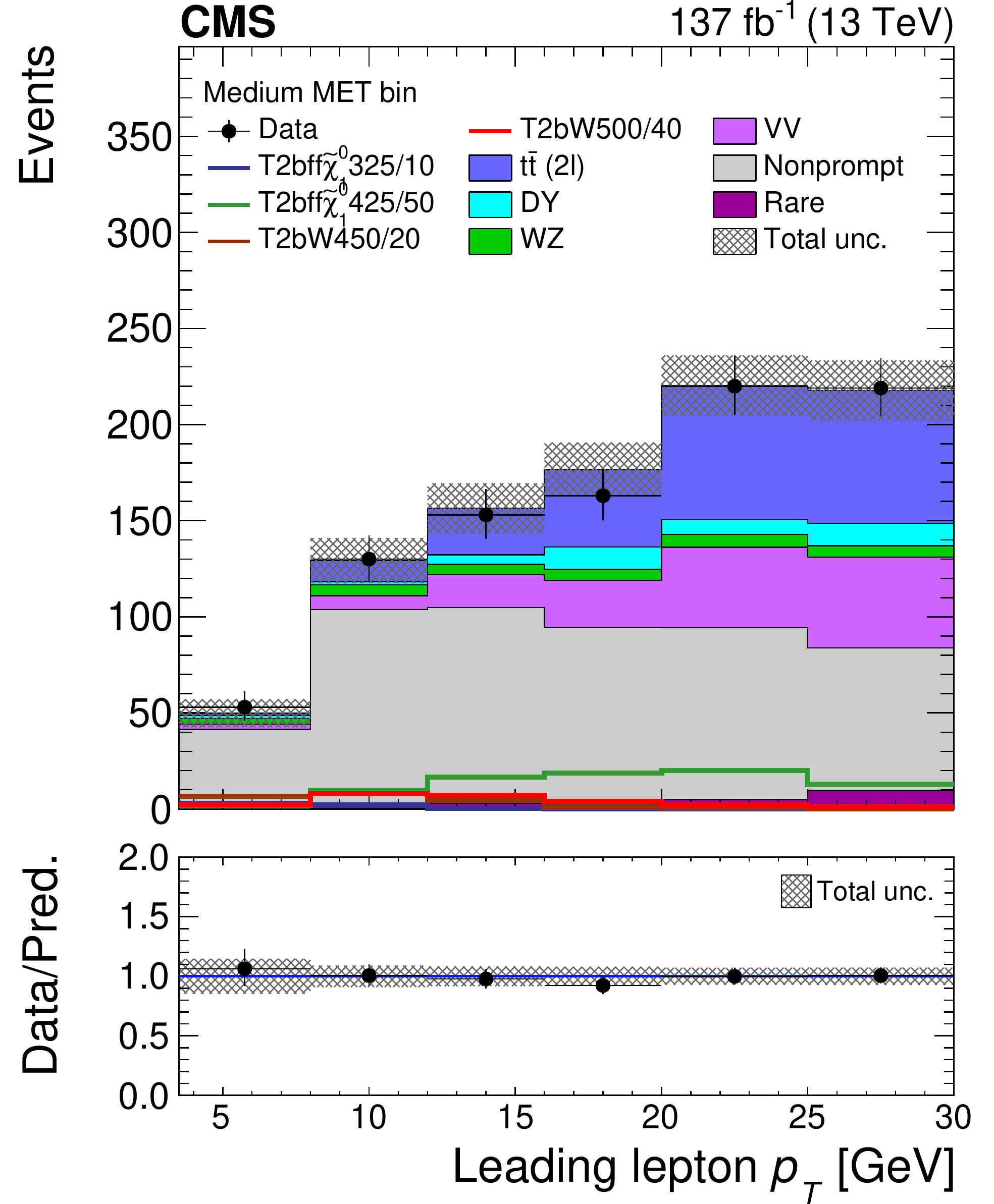}
    \includegraphics[width=0.49\textwidth]{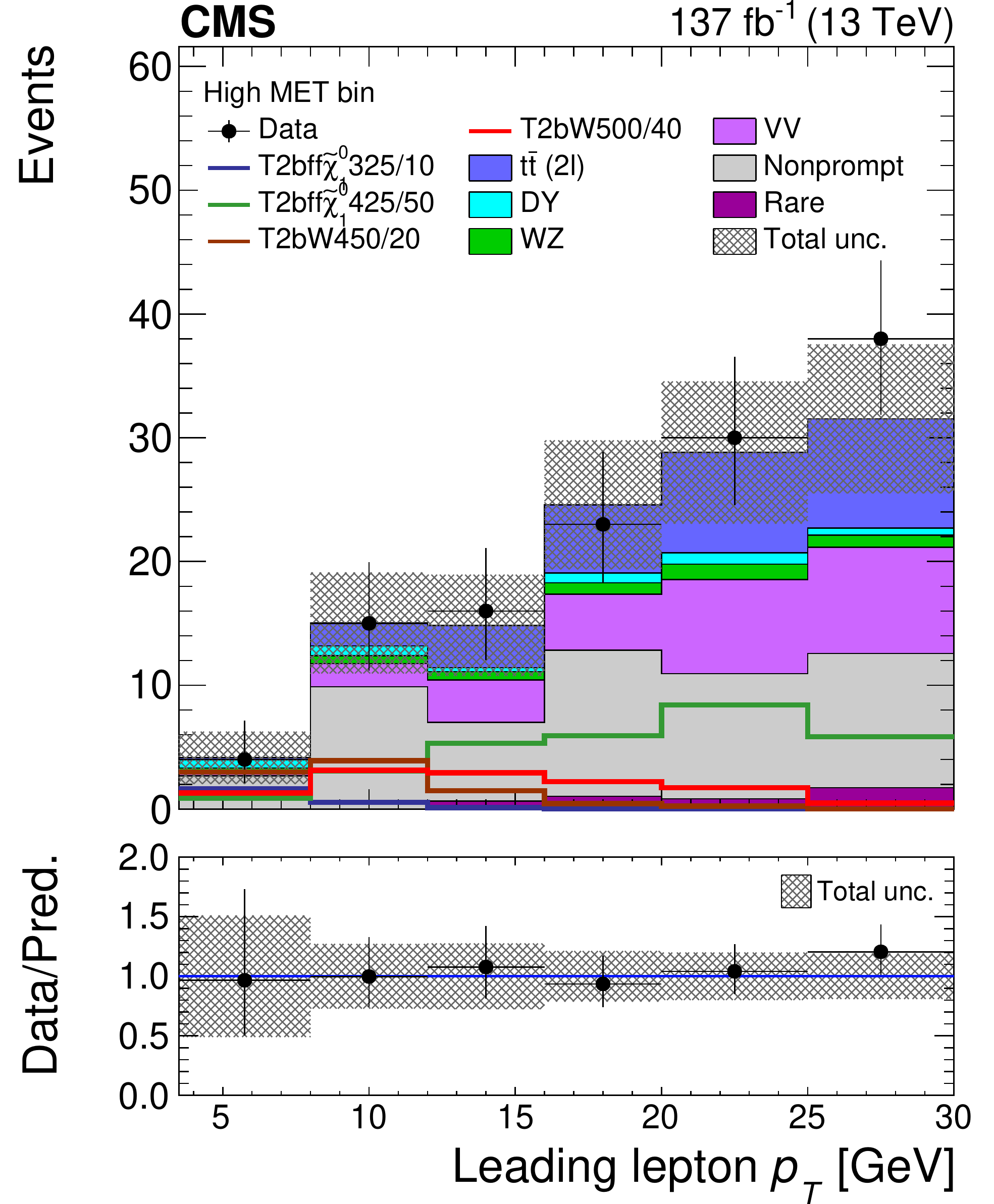}
    \includegraphics[width=0.49\textwidth]{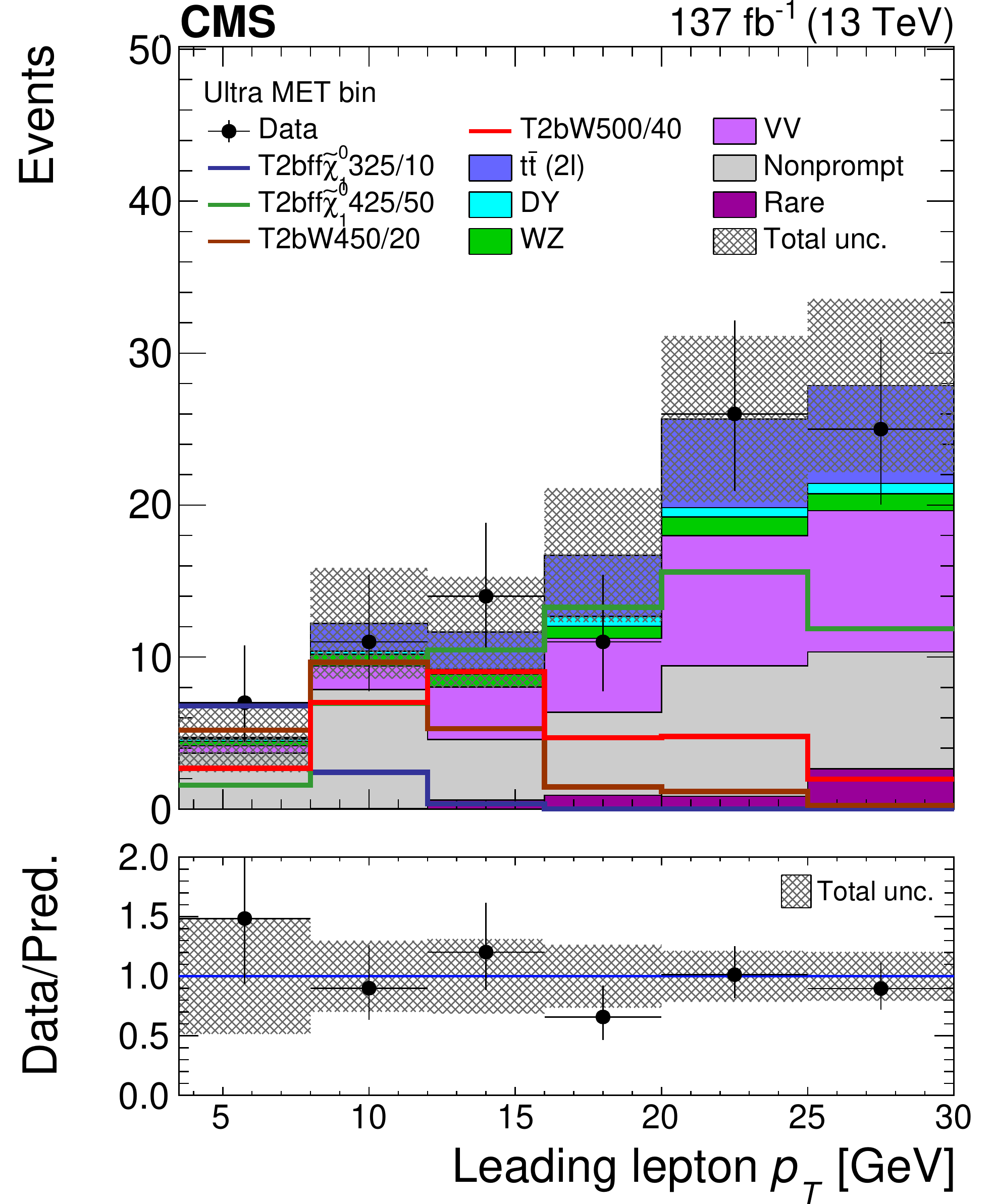}
\caption{The 2\Pell-Stop SR: the post-fit distribution of the leading lepton \pt variable is shown for the low- (upper left), medium- (upper right), high- (lower left) and ultra- (lower right) MET bins. Uncertainties include both the statistical and systematic components. The signal distributions overlaid on the plot are from the \Ttbffc and the \textsc{T2bW} models. The numbers after the model name in the legend indicate the mass of the top squark and the mass splitting between the top squark and LSP, in {\GeVns}.}
\label{fig:results:2lColSR}
\end{figure}

The estimated yields for the SM background processes and the data observed in the 2\Pell- and 3\Pell-Ewk SRs are shown in Figs.~\ref{fig:results:2lEwkSR} and \ref{fig:results:3lSR} respectively, while Fig.~\ref{fig:results:2lColSR} shows the 2\Pell-Stop SRs. The estimated yields from the different backgrounds and the data are also summarized in Tables \ref{tab:results:2lEwkSR}--\ref{tab:results:2lColSR} for each SR. The estimates correspond to the post-fit results (background-only), extracted from the maximum likelihood fit to the data. The uncertainties in the predicted yields include both statistical and systematic components, as described in Section~\ref{sec:systematics}. No significant deviation from the SM prediction is observed in the data.

\begin{table}[!hbt]
\centering
\topcaption{Observed and predicted yields as extracted from the maximum likelihood fit, in the 2\Pell-Ewk SRs. Uncertainties include both the statistical and systematic components.}
\label{tab:results:2lEwkSR}
\cmsTable{
\renewcommand{\arraystretch}{1.3}
\begin{tabular}{ccccccccccc}
\hline
\ptmiss [{\GeVns}] & \mll [{\GeVns}] & \ttbar & DY & VV & $\PW\PZ$ & Rare & Nonprompt & Total bkg & Data \\
\hline
\multirow{4}{*}{125--200}
& 4--10  & $4.0 \pm 2.0$  & $20.6 \pm 5.2$ & $3.7 \pm 2.4$ & $8.3 \pm 2.6$ & $0.28^{+0.72}_{-0.27}$ & $31.9 \pm 5.6$ & $68.7 \pm 8.7$  & $73$  \\
& 10--20 & $16.5 \pm 4.2$ & $28.0 \pm 6.2$ & $6.2 \pm 3.2$ & $6.5 \pm 2.3$ & $2.8 \pm 2.1$          & $90.1 \pm 9.3$  & $151 \pm 13$   & $165$ \\
& 20--30 & $18.0 \pm 4.4$ & $36.3 \pm 7.1$ & $7.8 \pm 3.5$ & $3.5 \pm 1.7$ & $2.9 \pm 2.1$          & $82.1 \pm 8.9$ & $151 \pm 13$   & $156$ \\
& 30--50 & $22.4 \pm 4.9$ & $10.2 \pm 3.7$ & $7.4 \pm 3.5$ & $1.3 \pm 1.0$ & $2.1 \pm 1.8$          & $39.6 \pm 6.2$ & $82.9 \pm 9.6$ & $80$ \\[\cmsTabSkip]

\multirow{5}{*}{200--240}
& 1--4   & $0.11^{+0.33}_{-0.10}$ & $0.37^{+0.72}_{-0.36}$ & $0.7^{+1.1}_{-0.7}$ & $1.3 \pm 1.0$ & $0.04^{+0.23}_{-0.03}$ & $3.0 \pm 2.0$  & $5.5 \pm 2.5$ & $2$  \\
& 4--10  & $0.75^{+0.90}_{-0.74}$ & $0.15^{+0.50}_{-0.14}$ & $1.4^{+1.5}_{-1.4}$                          & $3.5 \pm 1.7$   & $0.14^{+0.39}_{-0.13}$ & $11.9 \pm 3.6$ & $17.8 \pm 4.4$ & $19$ \\
& 10--20 & $2.9 \pm 1.7$          & $7.9 \pm 3.4$          & $2.9 \pm 2.2$                          & $2.5 \pm 1.4$   & $1.2 \pm 1.2$        & $42.8 \pm 6.8$ & $60.1 \pm 8.3$ & $59$ \\
& 20--30 & $4.3 \pm 2.1$          & $4.7 \pm 2.6$          & $2.6 \pm 2.0$                          & $1.1 \pm 1.0$ & $0.27^{+0.54}_{-0.26}$    & $31.3 \pm 5.8$ & $44.3 \pm 7.1$ & $47$ \\
& 30--50 & $5.7 \pm 2.4$          & $0.6^{+1.0}_{-0.6}$ & $2.8 \pm 2.1$                          & $0.63^{+0.70}_{-0.62}$ & $0.35^{+0.65}_{-0.34}$ & $17.6 \pm 4.4$ & $27.7 \pm 5.6$ & $24$ \\[\cmsTabSkip]

\multirow{5}{*}{240--290}
& 1--4   & $<0.02$                & $<0.1$                 & $0.43^{+0.88}_{-0.42}$ & $0.8 \pm 0.8$        & $<0.07$                & $1.5 \pm 1.3$  & $2.7 \pm 1.9$  & $2$  \\
& 4--10  & $0.9^{+1.2}_{-0.9}$ & $0.57^{+0.97}_{-0.56}$    & $0.8^{+1.1}_{-0.8}$    & $1.5 \pm 1.1$          & $0.3^{+2.6}_{-0.3}$ & $3.7 \pm 2.0$  & $7.7 \pm 3.9$  & $11$ \\
& 10--20 & $2.4 \pm 1.6$          & $3.4 \pm 2.3$          & $1.6 \pm 1.6$                          & $1.2 \pm 0.9$          & $0.3^{+1.3}_{-0.3}$ & $14.9 \pm 4.0$ & $23.8 \pm 5.4$ & $19$ \\
& 20--30 & $2.0 \pm 1.5$          & $2.4 \pm 1.9$          & $1.9 \pm 1.7$                          & $0.61^{+0.67}_{-0.60}$ & $0.03^{+0.45}_{-0.02}$ & $10.1 \pm 3.3$ & $17.0 \pm 4.5$ & $13$ \\
& 30--50 & $2.3 \pm 1.7$          & $0.32^{+0.73}_{-0.31}$ & $1.2^{+1.4}_{-1.1}$    & $0.40^{+0.53}_{-0.39}$ & $0.8^{+4.6}_{-0.7}$    & $6.6 \pm 2.7$  & $11.6 \pm 5.8$ & $10$ \\[\cmsTabSkip]

\multirow{5}{*}{$>290$}
& 1--4   & $<0.02$                & $<0.1$                 & $0.18^{+0.65}_{-0.17}$ & $0.57^{+0.65}_{-0.56}$ & $<0.01$                & $0.70^{+0.88}_{-0.69}$ & $1.5 \pm 1.3$  & $1$  \\
& 4--10  & $0.38^{+0.64}_{-0.37}$ & $0.8^{+1.1}_{-0.8}$    & $0.9^{+1.2}_{-0.9}$    & $1.3 \pm 1.0$        & $0.12^{+0.44}_{-0.11}$ & $1.7 \pm 1.3$          & $5.2 \pm 2.5$  & $3$  \\
& 10--20 & $1.3 \pm 1.2$          & $0.8^{+1.2}_{-0.8}$    & $1.6 \pm 1.6$                          & $1.05 \pm 0.89$        & $0.9^{+1.4}_{-0.9}$    & $7.8 \pm 2.9$          & $13.5 \pm 4.1$ & $15$ \\
& 20--30 & $0.9^{+1.0}_{-0.9}$ & $0.06^{+0.28}_{-0.05}$ & $1.5^{+1.6}_{-1.5}$                      & $0.3^{+0.50}_{-0.34}$ & $<0.09$                & $5.9 \pm 2.5$          & $8.8 \pm 3.2$  & $13$ \\
& 30--50 & $1.2 \pm 1.1$         & $<0.1$                 & $1.3^{+1.5}_{-1.3}$    & $0.09^{+0.24}_{-0.08}$ & $0.7^{+1.2}_{-0.7}$    & $3.6 \pm 2.0$          & $6.8 \pm 3.0$  & $9$  \\
\hline
\end{tabular}
}
\end{table}

\begin{table}[!hbt]
\centering
\topcaption{Observed and predicted yields as extracted from the maximum likelihood fit, in the 3\Pell-Ewk SRs. Uncertainties include both the statistical and systematic components.}
\label{tab:results:3lSR}
\cmsTable{
\renewcommand{\arraystretch}{1.3}
\begin{tabular}{ccccccccccc}
\hline
\ptmiss [{\GeVns}] & \mmnsfosll [{\GeVns}] & VV & $\PW\PZ$ & Rare & Nonprompt & Total bkg & Data \\

\hline
\multirow{4}{*}{125--200}
& 4--10  & $0.18^{+0.54}_{-0.17}$ & $4.8 \pm 1.9$          & $0.08^{+0.38}_{-0.07}$ & $0.61^{+0.83}_{-0.60}$ & $5.7 \pm 2.2$ & $3$ \\
& 10--20 & $0.08^{+0.35}_{-0.07}$ & $2.3 \pm 1.3$          & $0.5^{+1.0}_{-0.5}$ & $1.9 \pm 1.4$          & $4.9 \pm 2.2$ & $7$ \\
& 20--30 & $0.03^{+0.23}_{-0.02}$ & $1.0 \pm 1.0$        & $0.07^{+0.35}_{-0.06}$ & $1.3 \pm 1.2$          & $2.4 \pm 1.5$ & $4$ \\
& 30--50 & $0.01^{+0.13}_{-0.01}$ & $0.39^{+0.55}_{-0.38}$ & $0.08^{+0.37}_{-0.07}$ & $1.4 \pm 1.2$          & $1.8 \pm 1.4$ & $1$ \\[\cmsTabSkip]

\multirow{5}{*}{$>200$}
& 1--4   & $0.01^{+0.18}_{-0.01}$ & $1.5 \pm 1.0$          & $0.03^{+0.20}_{-0.02}$ & $0.18^{+0.44}_{-0.17}$ & $1.7 \pm 1.2$ & $3$ \\
& 4--10  & $0.05^{+0.34}_{-0.04}$ & $2.9 \pm 1.4$          & $0.16^{+0.47}_{-0.15}$ & $0.85^{+0.99}_{-0.84}$ & $4.0 \pm 1.8$ & $1$ \\
& 10--20 & $0.06^{+0.32}_{-0.05}$ & $2.0 \pm 1.2$          & $0.05^{+0.26}_{-0.04}$ & $2.1 \pm 1.5$          & $4.2 \pm 2.0$ & $5$ \\
& 20--30 & $<0.002$               & $0.52^{+0.60}_{-0.51}$ & $0.06^{+0.29}_{-0.05}$ & $1.1 \pm 1.1$          & $1.7 \pm 1.3$ & $2$ \\
& 30--50 & $<0.002$               & $0.31^{+0.46}_{-0.30}$ & $0.03^{+0.23}_{-0.02}$ & $1.0 \pm 1.0$ & $1.3 \pm 1.1$ & $1$ \\
\hline
\end{tabular}
}
\end{table}

\newpage

\begin{table}[!hbt]
\centering
\topcaption{Observed and predicted yields as extracted from the maximum likelihood fit, in the $\PW\PZ$-like selection SRs. Uncertainties include both the statistical and systematic components.}
\label{tab:results:WZSR}
\cmsTable{
\renewcommand{\arraystretch}{1.3}
\begin{tabular}{ccccccccccc}
\hline
\ptmiss [{\GeVns}] & \mmnsfosll [{\GeVns}] & VV & $\PW\PZ$ & Rare & Nonprompt & Total bkg & Data \\

\hline
\multirow{3}{*}{125--200}
& 4--10  & $0.13^{+0.47}_{-0.12}$ & $2.6 \pm 1.4$          & $0.31^{+0.67}_{-0.30}$ & $0.49^{+0.70}_{-0.48}$ & $3.5 \pm 1.8$ & $4$ \\
& 10--20 & $0.14^{+0.47}_{-0.13}$ & $4.3 \pm 1.8$          & $0.47^{+0.83}_{-0.46}$ & $1.2 \pm 1.1$          & $6.1 \pm 2.3$ & $11$ \\
& 20--30 & $0.17^{+0.51}_{-0.16}$ & $5.0 \pm 2.0$        & $0.50^{+0.85}_{-0.49}$ & $2.1 \pm 1.5$          & $7.8 \pm 2.6$ & $9$ \\[\cmsTabSkip]

\multirow{4}{*}{$>200$}
& 1--4   & $0.16^{+0.56}_{-0.15}$ & $0.11^{+0.29}_{-0.10}$ & $0.06^{+0.33}_{-0.05}$ & $0.44^{+0.66}_{-0.43}$ & $0.78^{+0.97}_{-0.77}$ & $0$ \\
& 4--10  & $0.22^{+0.60}_{-0.21}$ & $2.6 \pm 1.4$          & $0.10^{+0.38}_{-0.09}$ & $0.24^{+0.59}_{-0.23}$ & $3.1 \pm 1.6$          & $3$ \\
& 10--20 & $0.7^{+1.1}_{-0.7}$ & $10.6 \pm 2.8$         & $0.9^{+1.1}_{-0.9}$ & $1.9 \pm 1.4$             & $14.0 \pm 3.4$         & $19$ \\
& 20--30 & $0.7^{+1.0}_{-0.7}$ & $15.2 \pm 3.3$         & $1.2^{+1.3}_{-1.2}$ & $4.0 \pm 2.0$             & $21.0 \pm 4.2$         & $23$ \\
\hline
\end{tabular}
}
\end{table}

\begin{table}[!hbtp]
\centering
\topcaption{Observed and predicted yields as extracted from the maximum likelihood fit, in the 2\Pell-Stop SRs. Uncertainties include both the statistical and systematic components.}
\label{tab:results:2lColSR}
\cmsTable{
\renewcommand{\arraystretch}{1.3}
\begin{tabular}{ccccccccccc}
\hline
\ptmiss [{\GeVns}] & $\pt(\Pell_1)$ [{\GeVns}] & \ttbar & DY & VV & $\PW\PZ$ & Rare & Nonprompt & Total bkg & Data \\

\hline
\multirow{6}{*}{125--200}
& 3.5--8 & $1.2 \pm 1.2$  & $5.2 \pm 3.1$  & $1.0^{+1.2}_{-1.0}$ & $1.4 \pm 1.1$ & $0.06^{+0.27}_{-0.05}$ & $41.0 \pm 6.3$ & $49.9 \pm 7.2$ & $52$  \\
& 8--12  & $15.0 \pm 4.0$ & $22.9 \pm 5.9$ & $6.6 \pm 3.1$ & $6.0 \pm 2.1$ & $0.96^{+0.99}_{-0.95}$ & $93.1 \pm 9.4$   & $144 \pm 12$   & $156$ \\
& 12--16 & $31.8 \pm 5.9$ & $24.0 \pm 6.1$ & $13.7 \pm 4.5$ & $7.2 \pm 2.4$ & $2.8 \pm 1.7$ & $101.3 \pm 9.9$   & $180 \pm 14$   & $196$ \\
& 16--20 & $59.9 \pm 8.0$ & $36.9 \pm 7.5$ & $19.8 \pm 5.5$ & $7.9 \pm 2.5$ & $4.2 \pm 2.1$ & $100.2 \pm 9.8$   & $229 \pm 16$  & $238$ \\
& 20--25 & $103 \pm 11$ & $27.2 \pm 6.5$ & $33.2 \pm 7.1$ & $7.7 \pm 2.5$ & $7.5 \pm 2.8$ & $95.0 \pm 9.5$   & $273 \pm 18$  & $285$ \\
& 25--30 & $114 \pm 11$ & $21.4 \pm 5.7$ & $35.5 \pm 7.3$ & $5.1 \pm 2.0$ & $8.0 \pm 2.8$ & $71.5 \pm 8.3$ & $256 \pm 17$  & $246$ \\[\cmsTabSkip]

\multirow{6}{*}{200--290}
& 3.5--8 & $1.1 \pm 1.0$ & $1.7 \pm 1.5$  & $2.8 \pm 2.1$  & $2.9 \pm 1.4$ & $0.04^{+0.20}_{-0.03}$ & $41.3 \pm 6.6$ & $49.9 \pm 7.3$ & $53$  \\
& 8--12  & $11.0 \pm 3.3$   & $1.6 \pm 1.5$  & $7.3 \pm 3.3$  & $5.6 \pm 2.0$ & $0.43^{+0.65}_{-0.42}$ & $103 \pm 10$   & $129 \pm 12$   & $130$ \\
& 12--16 & $24.1 \pm 4.9$  & $5.0 \pm 2.6$  & $17.1 \pm 5.0$ & $5.5 \pm 2.0$ & $2.9 \pm 1.7$          & $102 \pm 10$   & $156 \pm 13$   & $153$ \\
& 16--20 & $40.3 \pm 6.3$  & $11.7 \pm 4.2$ & $24.7 \pm 6.1$ & $5.6 \pm 2.0$ & $2.4 \pm 1.6$          & $92.0 \pm 9.8$ & $177 \pm 14$   & $163$ \\
& 20--25 & $69.9 \pm 8.3$  & $7.6 \pm 3.4$  & $41.9 \pm 7.9$ & $6.7 \pm 2.2$ & $5.0 \pm 2.2$          & $89.3 \pm 9.7$ & $220 \pm 16$   & $220$ \\
& 25--30 & $69.0 \pm 8.3$  & $11.8 \pm 4.1$ & $47.3 \pm 8.4$ & $5.9 \pm 2.0$ & $9.6 \pm 3.1$          & $74.2 \pm 8.9$ & $218 \pm 16$   & $219$ \\[\cmsTabSkip]

\multirow{6}{*}
{290--340}
& 3.5--8 & $0.15^{+0.35}_{-0.14}$       & $0.67^{+0.90}_{-0.66}$ & $0.34^{+0.72}_{-0.33}$ & $0.29^{+0.44}_{-0.28}$ & $<0.05$ & $2.7 \pm 1.7$  & $4.1  \pm 2.1$ & $4$  \\
& 8--12  & $1.9 \pm 1.4$ & $0.8^{+1.1}_{-0.8}$ & $1.9 \pm 1.7$ & $0.64^{+0.67}_{-0.63}$ & $0.01^{+0.11}_{-0.01}$ & $9.9 \pm 3.2$  & $15.0 \pm 4.1$ & $15$ \\
& 12--16 & $3.4 \pm 1.8$ & $0.33^{+0.61}_{-0.32}$ & $3.4 \pm 2.3$ & $0.69 \pm 0.69$ & $0.64^{+0.85}_{-0.63}$ & $6.4 \pm 2.6$  & $14.8 \pm 4.1$ & $16$ \\
& 16--20 & $5.5 \pm 2.3$ & $0.8^{+1.1}_{-0.8}$ & $4.5 \pm 2.6$ & $0.91 \pm 0.80$ & $1.0 \pm 1.0$  & $11.8 \pm 3.5$ & $24.6 \pm 5.2$ & $23$ \\[0.5ex]
& 20--25 & $8.1 \pm 2.8$ & $0.9^{+1.2}_{-0.9}$    & $7.6 \pm 3.4$ & $1.24 \pm 0.93$ & $0.82^{+0.89}_{-0.81}$    & $10.1 \pm 3.2$ & $28.8 \pm 5.8$ & $30$ \\
& 25--30 & $8.8 \pm 2.9$ & $0.58^{+0.97}_{-0.57}$  &  $8.6 \pm 3.6$ & $0.96 \pm 0.81$ & $1.7 \pm 1.3$ & $10.8 \pm 3.4$ & $31.5 \pm 6.0$ & $38$ \\[\cmsTabSkip]

\multirow{6}{*}{$>340$}
& 3.5--8 & $0.12^{+0.37}_{-0.11}$ & $0.14^{+0.51}_{-0.13}$ & $0.48^{+0.86}_{-0.47}$ & $0.29^{+0.46}_{-0.28}$ & $<0.03$  & $3.7 \pm 2.0$ & $4.7 \pm 2.3$  & $7$  \\
& 8--12  & $1.8 \pm 1.3$ & $0.22^{+0.59}_{-0.21}$ & $1.5 \pm 1.5$  & $0.78 \pm 0.75$ & $0.02^{+0.12}_{-0.01}$ & $7.8 \pm 2.9$ & $12.2 \pm 3.6$ & $11$ \\
& 12--16 & $2.4 \pm 1.5$ & $0.31^{+0.63}_{-0.30}$ & $3.5 \pm 2.3$ & $0.87 \pm 0.78$  & $0.60^{+0.79}_{-0.59}$ & $4.0 \pm 2.0$ & $11.6 \pm 3.6$ & $14$ \\
& 16--20 & $4.0 \pm 2.0$ & $0.64^{+0.89}_{-0.63}$ & $4.9 \pm 2.7$ & $0.80 \pm 0.75$ & $0.90^{+0.93}_{-0.89}$ & $5.5 \pm 2.5$ & $16.7 \pm 4.4$ & $11$ \\
& 20--25 & $5.8 \pm 2.3$ & $0.62^{+0.95}_{-0.61}$ & $8.6 \pm 3.6$ & $1.22 \pm 0.93$  & $0.84^{+0.92}_{-0.83}$ & $8.6 \pm 3.0$ & $25.7 \pm 5.5$ & $26$ \\
& 25--30 & $6.5 \pm 2.5$ & $0.7^{+1.0}_{-0.7}$    & $9.3 \pm 3.7$ & $1.12 \pm 0.88$ & $2.6 \pm 1.6$ & $7.7 \pm 2.9$ & $27.9 \pm 5.7$ & $25$ \\[0.5ex]
\hline
\end{tabular}
}
\end{table}

\section{Interpretation}\label{sec:interpretations}

The results of the search are interpreted in the context of the SUSY scenarios described in Section~\ref{sec:sample}. Limits on the production cross section for SUSY particle pairs as a function of their masses are computed using a modified frequentist approach that employs the \CLs criterion and an asymptotic formulation~\cite{frequentist_limit,Read:2002hq,Cowan:2010js,LHC-HCG}.

\begin{figure}[!hbtp]
  \centering
    \includegraphics[width=0.80\textwidth]{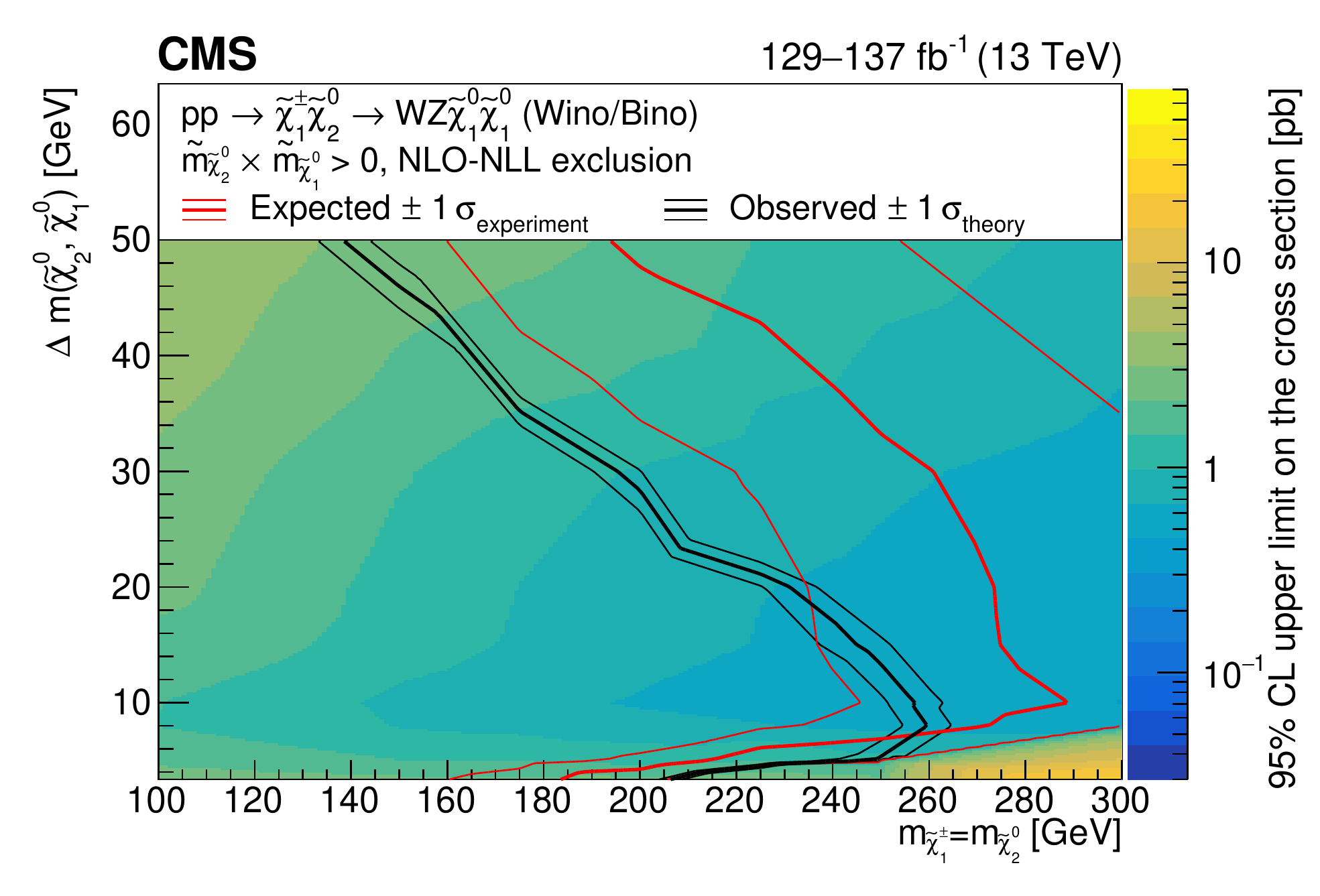}
    \includegraphics[width=0.80\textwidth]{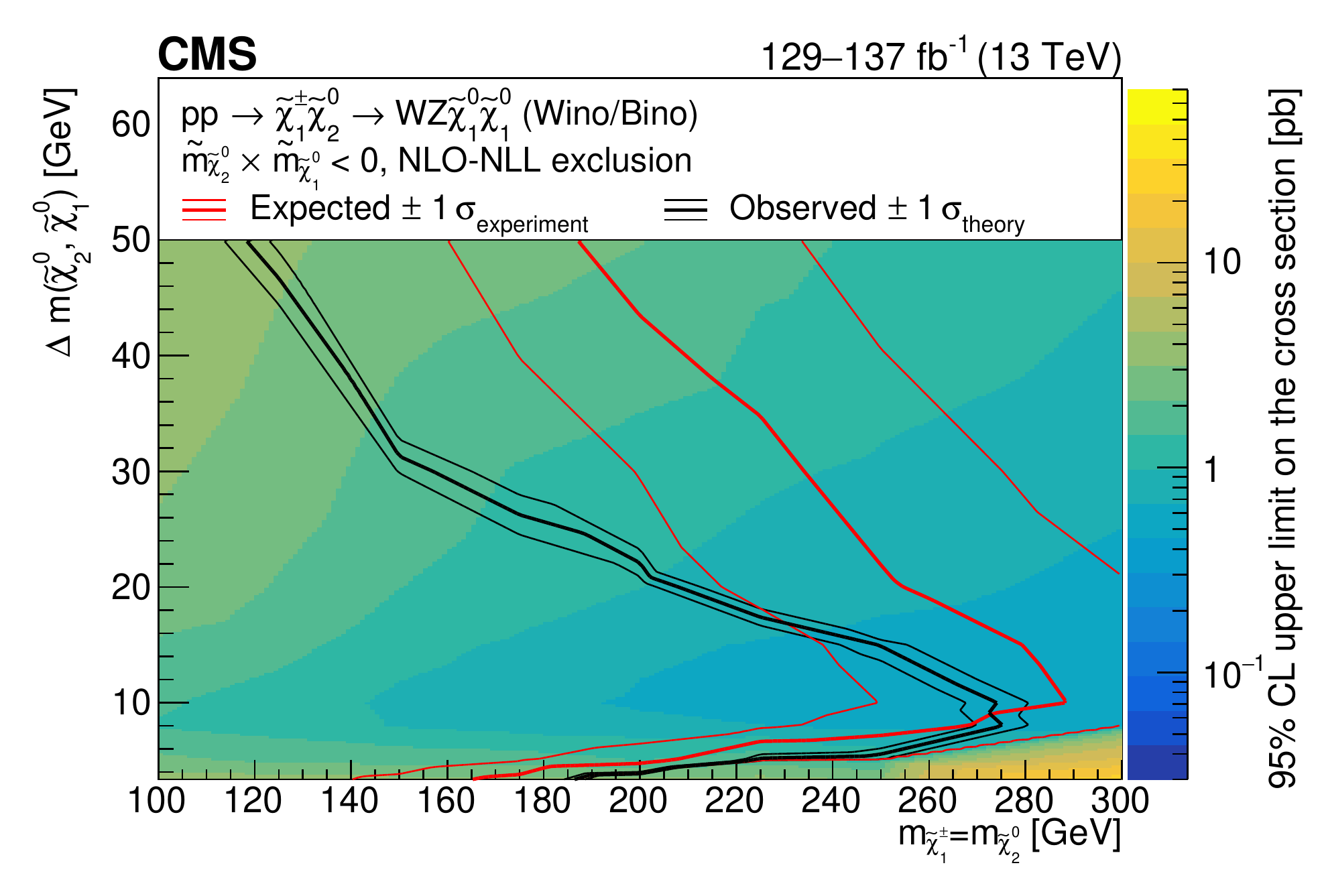}
  \caption{The observed 95\% \CL exclusion contours (black curves) assuming the NLO+NLL cross sections, with the variations (thin lines) corresponding to the uncertainty in the cross section for the \TChiWZ model. The red curves present the 95\% \CL expected limits with the band (thin lines) covering 68\% of the limits in the absence of signal. Results are reported for the $\mczt\mczo>0(<0)$ \mll spectrum reweighting scenario in the upper (lower) plot. The range of luminosities of the analysis regions included in the fit is indicated on the plot.}
  \label{fig:interpretations:tchiwz}
\end{figure}

We express the results as upper limits at 95\% confidence level (\CL) on the potential presence of a SUSY signal in the data. The limits of the \TChiWZ\ simplified model for the production of a wino pair decaying into bino LSPs are shown for the two signal scenarios described in Section~\ref{sec:sample}. When considering the full matrix element of the electroweakino decay, the relative sign of the eigenvalues of the neutralino mass matrix leads to two slightly different \mll distributions in \PSGczDt decays. The upper plot in Fig.~\ref{fig:interpretations:tchiwz} shows the limit for the scenario where the product of \mczo, \mczt eigenvalues is positive, while the negative case is shown by the lower plot. The results are reweighted to account for the branching fraction modulation of the off-shell \PW and \PZ bosons, as mentioned in Section~\ref{sec:sample}.

The observed exclusion limit shown in Fig.~\ref{fig:interpretations:tchiwz} is weaker than the expected one in the intermediate and high $\Delta m$ ranges. This is due to data yields that are higher than the predictions in specific bins of the 2\Pell-Ewk SR (Table~\ref{tab:results:2lEwkSR}, ultra-MET bin, $20 < \mll \leq 30\GeV$), the 3\Pell-Ewk SR (Table~\ref{tab:results:3lSR}, low-MET bin, $10 < \mmnsfosll \leq 30\GeV$) and $\PW\PZ$-like selection SR (Table~\ref{tab:results:WZSR}, low- and high-MET bin, $10 < \mmnsfosll \leq 20\GeV$). The shift of the \mll distribution for the negative case for the product of \mczo, \mczt eigenvalues causes the observed upper limit signal cross section to be higher, leading to a smaller range of excluded mass parameters in this scenario. The excess has a maximum local significance of 2.4 standard deviations for the signal mass point with $m_{\PSGczDt} = 125\GeV$ and $\Delta m = 40\GeV$.

\begin{figure}[!hbtp]
  \centering
    \includegraphics[width=0.80\textwidth]{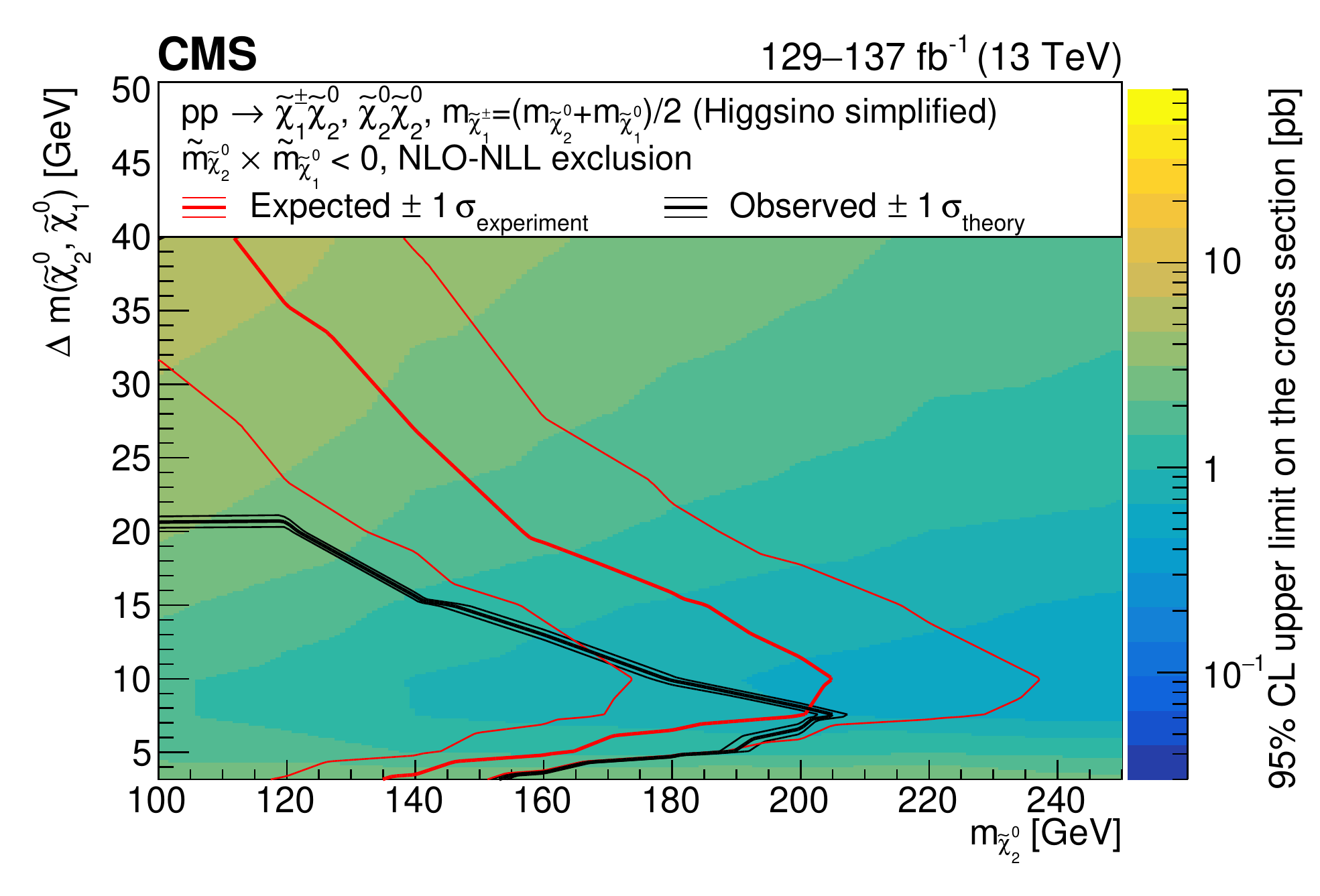}
    \includegraphics[width=0.80\textwidth]{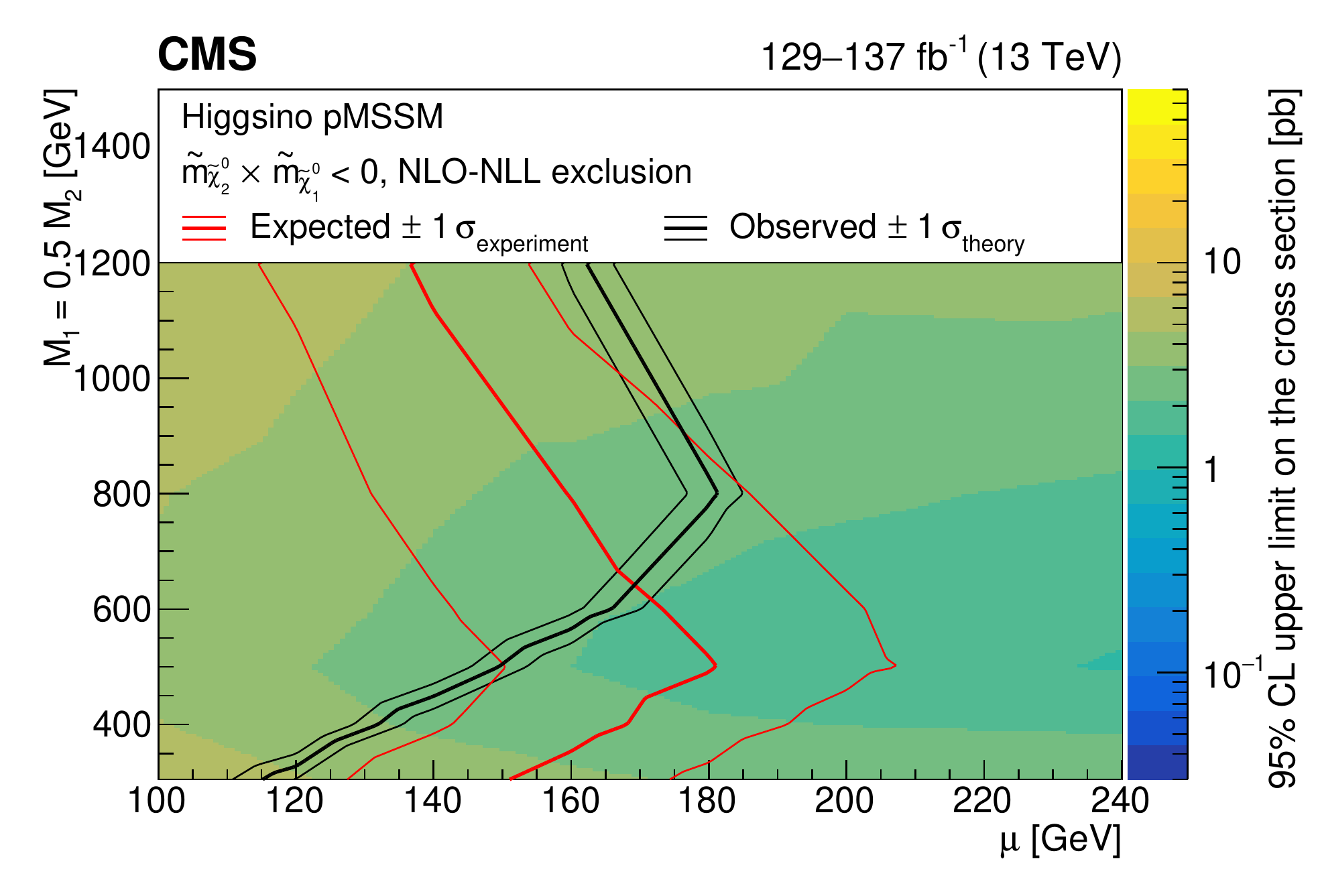}
  \caption{The observed 95\% \CL exclusion contours (black curves) assuming the NLO+NLL cross sections, with the variations (thin lines) corresponding to the uncertainty in the cross section for the simplified (upper) and the pMSSM (lower) higgsino models. The simplified model includes both neutralino pair and neutralino-chargino production modes, while the pMSSM one includes all possible production modes. The red curves present the 95\% \CL expected limits with the band (thin lines) covering 68\% of the limits in the absence of signal. The results are reported for the $\mczt\mczo<0$ \mll spectrum reweighting scenario. The range of luminosities of the analysis regions included in the fit is indicated on the plot.}
  \label{fig:interpretations:higgsino}
\end{figure}

The upper plot of Fig.~\ref{fig:interpretations:higgsino} displays the space of allowed electroweakino masses for the higgsino simplified model, assuming a chargino with mass halfway between the two lightest neutralinos. The simplified higgsino model includes neutralino pair production and neutralino-chargino production, while the pMSSM higgsino model includes all possible production modes. The dilepton invariant mass distributions are reweighted for the case where $\mczo\mczt < 0$, since this is the combination allowed when the higgsino is the LSP. The results also take into account the off-shell \PW and \PZ boson branching fraction corrections mentioned in Section~\ref{sec:sample}. A weaker than expected observed exclusion limit in the higher $\Delta m$ region is present also for this interpretation for the reasons mentioned above.

The expected and observed exclusion contours for the pMSSM higgsino model are shown in the lower plot of Fig.~\ref{fig:interpretations:higgsino}. The limits are presented in the plane of the higgsino-bino mass parameters $\mu$-$M_1$. In the pMSSM, larger $\mu$ values roughly correspond to larger masses for the parent SUSY particles. Larger values of the $M_1$ parameter correspond to smaller values of the mass difference between the LSP and its parent SUSY particle. Due to this, the fact that the observed limit is weaker than the expected one for intermediate and higher $\Delta m$ values of the \TChiWZ and higgsino simplified models manifests at small $M_1$ values for the pMSSM higgsino model.

\begin{figure}[!hbtp]
  \centering
    \includegraphics[width=0.80\textwidth]{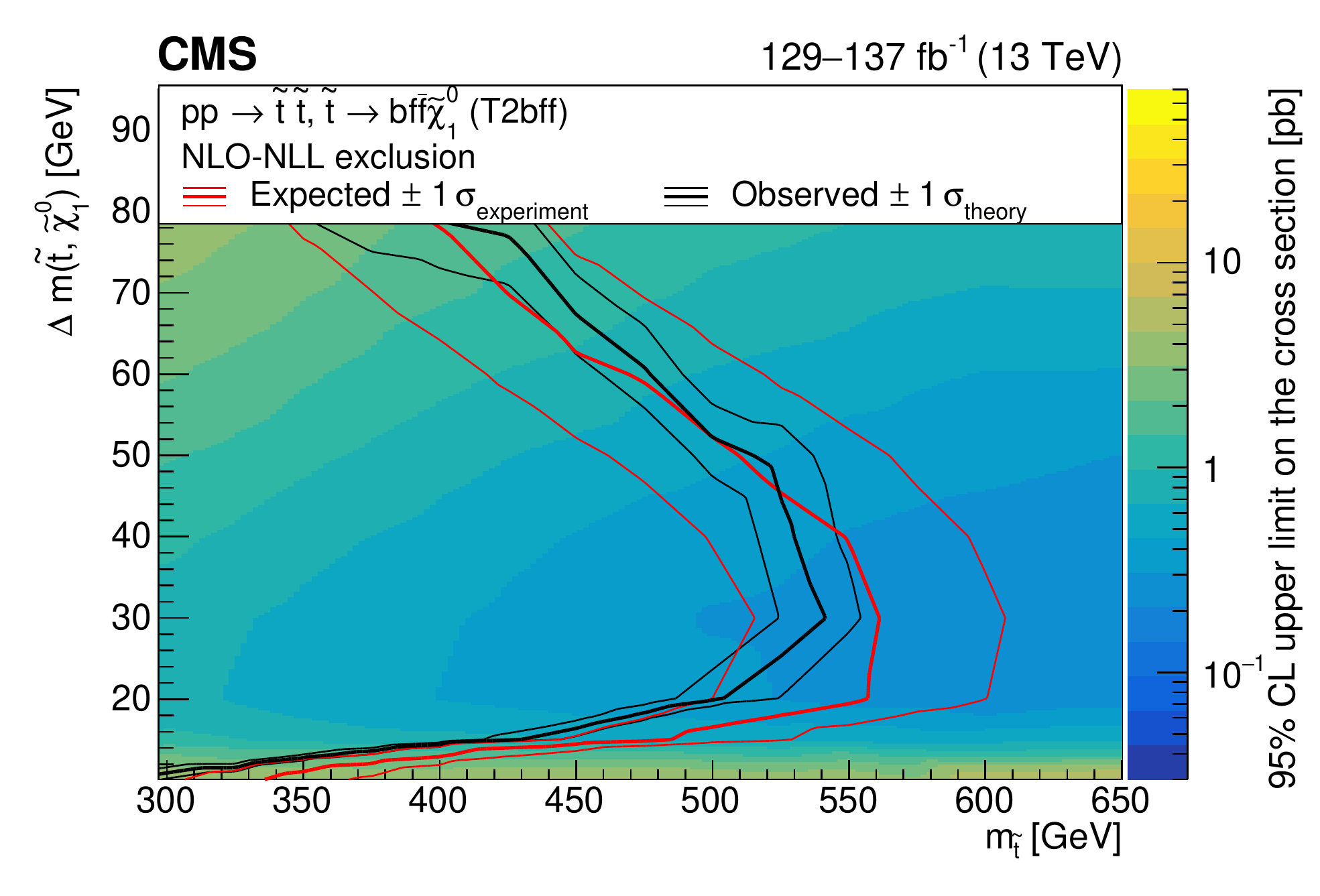}
    \includegraphics[width=0.80\textwidth]{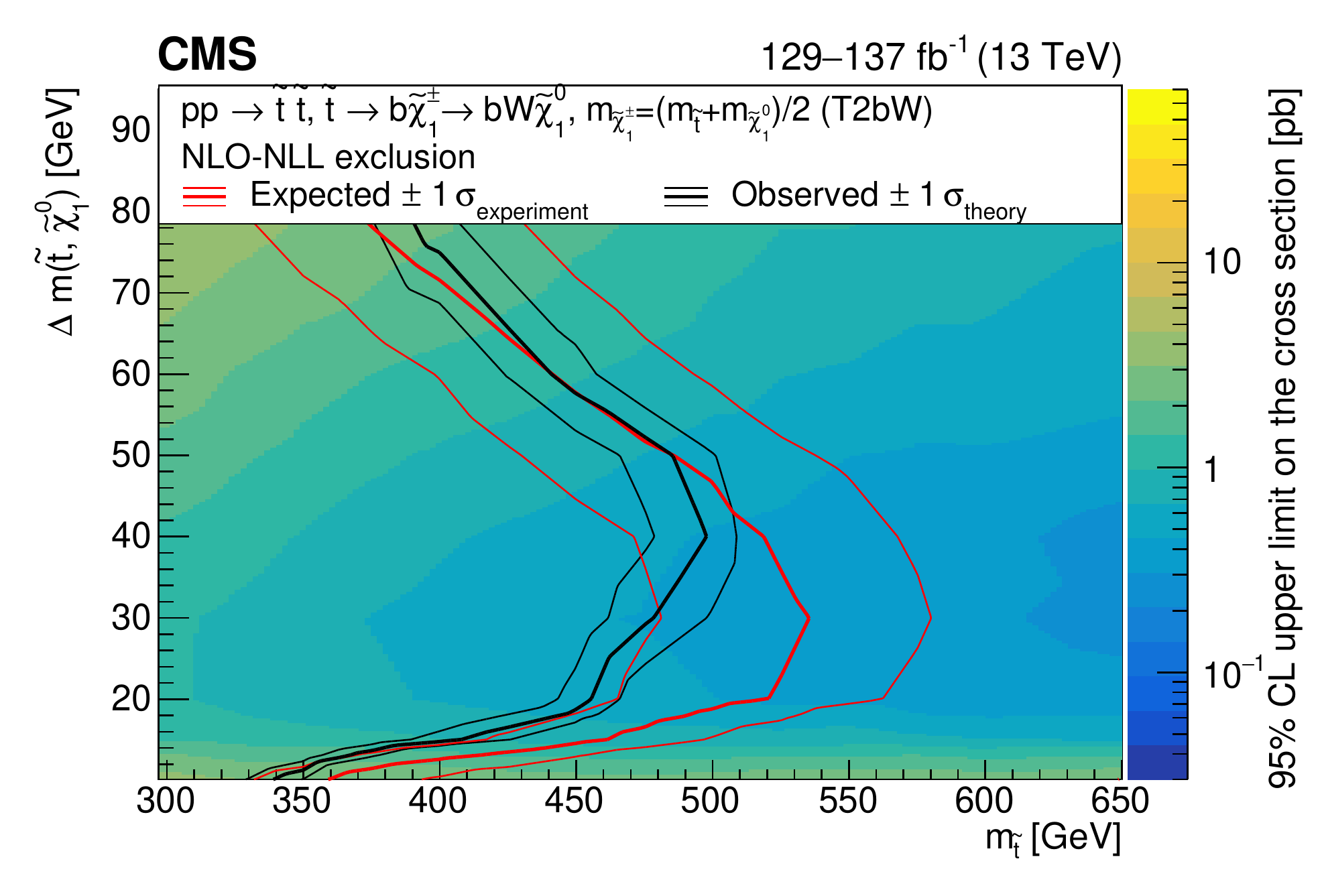}
  \caption{The observed 95\% \CL exclusion contours (black curves) assuming the NLO+NLL cross sections, with the variations (thin lines) corresponding to the uncertainty in the cross section for the \Ttbffc (upper) and \textsc{T2bW} (lower) simplified models. The red curves present the 95\% \CL expected limits with the band (thin lines) covering 68\% of the limits in the absence of signal. The range of luminosities of the analysis regions included in the fit is indicated on the plot.}
  \label{fig:interpretations:t2tt}
\end{figure}

For direct production of top squarks, the 2\Pell-Stop SRs are instead used in the maximum likelihood fit to extract upper limits on the \Ttbffc and \textsc{T2bW} models presented in Section~\ref{sec:sample}. Figure \ref{fig:interpretations:t2tt} shows the limits of the \Ttbffc (upper) and \textsc{T2bW} (lower) simplified models. For both models, the corrections for the modulation of the off-shell \PW bosons branching fraction are considered. The drop in the exclusion line around $\Delta m = 20\GeV$ is caused by the smaller acceptance when going to lower $\Delta m$, which is due to the minimum lepton \pt requirements of the SRs.

\clearpage

\section{Summary}\label{sec:summary}

A search for new physics is performed using events with two or three soft leptons and missing transverse momentum. These signatures are motivated by models predicting a weakly interacting massive particle that originates from the decay of another new particle with nearly degenerate mass. The results are based on data collected by the CMS experiment at the LHC during 2016--2018, corresponding to an integrated luminosity of up to 137\fbinv. The observed event yields are in agreement with the standard model expectations. 

The results are interpreted in the framework of supersymmetric (SUSY) simplified models targeting electroweakino mass-degenerate spectra and top squark-lightest neutralino (\sTop-$\PSGczDo$) mass-degenerate benchmark models. An interpretation of the analysis is performed also in the phenomenological minimal SUSY standard model (pMSSM) framework.
In particular, the simplified wino-bino model in which the next-to-lightest neutralino and the lightest chargino are produced and decay according to $\PSGczDt\PSGcpmDo\to\PZ^{*}\PW^{*}\PSGczDo\PSGczDo$ are explored for mass differences ($\Delta m$) between \PSGczDt and \PSGczDo of less than 50\GeV, assuming wino production cross sections. 
At 95\% confidence level, wino-like $\PSGcpmDo$/$\PSGczDt$ masses are excluded up to 275\GeV for $\Delta m$ of 10\GeV relative to the lightest neutralino.
The higgsino simplified model is of particular interest; mass-degenerate electroweakinos are expected in natural SUSY, which predicts light higgsinos. In this model, excluded masses reach up to 205\GeV for $\Delta m$ of 7.5\GeV and 150\GeV for a highly compressed scenario with $\Delta m$ of 3\GeV.
In the pMSSM higgsino model, the limits are presented in the plane of the higgsino-bino mass parameters $\mu$-$M_1$; the higgsino mass parameter $\mu$ is excluded up to 170\GeV, when the bino mass parameter $M_1$ is 600\GeV. For larger values of $M_1$, the mass splitting $\Delta m (\PSGczDt, \PSGczDo)$ becomes smaller; for $M_1 = 800\GeV$, $\mu$ is excluded up to 180\GeV.
Finally, two \sTop-$\PSGczDo$ mass-degenerate benchmark models are considered. Top squarks with masses below 540 (480)\GeV are excluded for the four-body (chargino-mediated) top squark decay model, with a (\sTop-$\PSGczDo$) mass splitting at 30\GeV.

\begin{acknowledgments}

  We congratulate our colleagues in the CERN accelerator departments for the excellent performance of the LHC and thank the technical and administrative staffs at CERN and at other CMS institutes for their contributions to the success of the CMS effort. In addition, we gratefully acknowledge the computing centers and personnel of the Worldwide LHC Computing Grid and other centers for delivering so effectively the computing infrastructure essential to our analyses. Finally, we acknowledge the enduring support for the construction and operation of the LHC, the CMS detector, and the supporting computing infrastructure provided by the following funding agencies: BMBWF and FWF (Austria); FNRS and FWO (Belgium); CNPq, CAPES, FAPERJ, FAPERGS, and FAPESP (Brazil); MES and BNSF (Bulgaria); CERN; CAS, MoST, and NSFC (China); MINCIENCIAS (Colombia); MSES and CSF (Croatia); RIF (Cyprus); SENESCYT (Ecuador); MoER, ERC PUT and ERDF (Estonia); Academy of Finland, MEC, and HIP (Finland); CEA and CNRS/IN2P3 (France); BMBF, DFG, and HGF (Germany); GSRI (Greece); NKFIA (Hungary); DAE and DST (India); IPM (Iran); SFI (Ireland); INFN (Italy); MSIP and NRF (Republic of Korea); MES (Latvia); LAS (Lithuania); MOE and UM (Malaysia); BUAP, CINVESTAV, CONACYT, LNS, SEP, and UASLP-FAI (Mexico); MOS (Montenegro); MBIE (New Zealand); PAEC (Pakistan); MSHE and NSC (Poland); FCT (Portugal); JINR (Dubna); MON, RosAtom, RAS, RFBR, and NRC KI (Russia); MESTD (Serbia); SEIDI, CPAN, PCTI, and FEDER (Spain); MOSTR (Sri Lanka); Swiss Funding Agencies (Switzerland); MST (Taipei); ThEPCenter, IPST, STAR, and NSTDA (Thailand); TUBITAK and TAEK (Turkey); NASU (Ukraine); STFC (United Kingdom); DOE and NSF (USA).
  
  \hyphenation{Rachada-pisek} Individuals have received support from the Marie-Curie program and the European Research Council and Horizon 2020 Grant, contract Nos.\ 675440, 724704, 752730, 758316, 765710, 824093, 884104, and COST Action CA16108 (European Union); the Leventis Foundation; the Alfred P.\ Sloan Foundation; the Alexander von Humboldt Foundation; the Belgian Federal Science Policy Office; the Fonds pour la Formation \`a la Recherche dans l'Industrie et dans l'Agriculture (FRIA-Belgium); the Agentschap voor Innovatie door Wetenschap en Technologie (IWT-Belgium); the F.R.S.-FNRS and FWO (Belgium) under the ``Excellence of Science -- EOS" -- be.h project n.\ 30820817; the Beijing Municipal Science \& Technology Commission, No. Z191100007219010; the Ministry of Education, Youth and Sports (MEYS) of the Czech Republic; the Deutsche Forschungsgemeinschaft (DFG), under Germany's Excellence Strategy -- EXC 2121 ``Quantum Universe" -- 390833306, and under project number 400140256 - GRK2497; the Lend\"ulet (``Momentum") Program and the J\'anos Bolyai Research Scholarship of the Hungarian Academy of Sciences, the New National Excellence Program \'UNKP, the NKFIA research grants 123842, 123959, 124845, 124850, 125105, 128713, 128786, and 129058 (Hungary); the Council of Science and Industrial Research, India; the Latvian Council of Science; the Ministry of Science and Higher Education and the National Science Center, contracts Opus 2014/15/B/ST2/03998 and 2015/19/B/ST2/02861 (Poland); the Funda\c{c}\~ao para a Ci\^encia e a Tecnologia, grant CEECIND/01334/2018 (Portugal); the National Priorities Research Program by Qatar National Research Fund; the Ministry of Science and Higher Education, projects no. 14.W03.31.0026 and no. FSWW-2020-0008, and the Russian Foundation for Basic Research, project No.19-42-703014 (Russia); the Programa Estatal de Fomento de la Investigaci{\'o}n Cient{\'i}fica y T{\'e}cnica de Excelencia Mar\'{\i}a de Maeztu, grant MDM-2015-0509 and the Programa Severo Ochoa del Principado de Asturias; the Hellenic Foundation for Research and Innovation (HFRI) (Project Number: 2288) and the Stavros Niarchos Foundation (Greece); the Rachadapisek Sompot Fund for Postdoctoral Fellowship, Chulalongkorn University and the Chulalongkorn Academic into Its 2nd Century Project Advancement Project (Thailand); the Kavli Foundation; the Nvidia Corporation; the SuperMicro Corporation; the Welch Foundation, contract C-1845; and the Weston Havens Foundation (USA).
\end{acknowledgments}

\bibliography{auto_generated}

\cleardoublepage \appendix\section{The CMS Collaboration \label{app:collab}}\begin{sloppypar}\hyphenpenalty=5000\widowpenalty=500\clubpenalty=5000\cmsinstitute{Yerevan~Physics~Institute, Yerevan, Armenia}
A.~Tumasyan
\cmsinstitute{Institut~f\"{u}r~Hochenergiephysik, Vienna, Austria}
W.~Adam\cmsorcid{0000-0001-9099-4341}, J.W.~Andrejkovic, T.~Bergauer\cmsorcid{0000-0002-5786-0293}, S.~Chatterjee\cmsorcid{0000-0003-2660-0349}, M.~Dragicevic\cmsorcid{0000-0003-1967-6783}, A.~Escalante~Del~Valle\cmsorcid{0000-0002-9702-6359}, R.~Fr\"{u}hwirth\cmsAuthorMark{1}, M.~Jeitler\cmsAuthorMark{1}\cmsorcid{0000-0002-5141-9560}, N.~Krammer, L.~Lechner\cmsorcid{0000-0002-3065-1141}, D.~Liko, I.~Mikulec, P.~Paulitsch, F.M.~Pitters, J.~Schieck\cmsAuthorMark{1}\cmsorcid{0000-0002-1058-8093}, R.~Sch\"{o}fbeck\cmsorcid{0000-0002-2332-8784}, M.~Spanring\cmsorcid{0000-0001-6328-7887}, S.~Templ\cmsorcid{0000-0003-3137-5692}, W.~Waltenberger\cmsorcid{0000-0002-6215-7228}, C.-E.~Wulz\cmsAuthorMark{1}\cmsorcid{0000-0001-9226-5812}
% ...
\vskip\cmsinstskip
% ...
1:~Also at TU~Wien, Wien, Austria\\
% ...
\end{sloppypar}
\end{document}